\documentclass[final,5p,times,twocolumn]{elsarticle}
 \biboptions{comma,sort&compress}
\usepackage{graphicx}
\usepackage{here}
\usepackage[dvips]{epsfig}
\def\beq{\begin{equation}}
\def\eeq{\end{equation}}
\def\bea{\begin{eqnarray}}
\def\eea{\end{eqnarray}}
\def\bq{\begin{quote}}
\def\eq{\end{quote}}

\def\nnb{\nonumber}
\def\ga{\left(}
\def\dr{\right)}

\def\lrar{\Longrightarrow}
		
\def\nnb{\nonumber}
\def\la{\langle}
\def\ra{\rangle}
\def\nin{\noindent}
\def\ba{\vspace*{-0.2cm}\begin{array}}
\def\ea{\end{array}\vspace*{-0.2cm}}

\def\b{$\bullet~$}
\def\als{\alpha_s}

\def\gg2{ \la\alpha_s G^2 \ra}
\def\gg3{g^3f_{abc}\la G^aG^bG^c \ra}
\def\ggg4{\la\als^2G^4\ra}

\def\beq{\begin{equation}}
\def\enq{\end{equation}}
\def\beqa{\begin{eqnarray}}
\def\enqa{\end{eqnarray}}
\def\nnb{\nonumber}

\def\qq{\lag\bar{q}q\rag}

\def\mix{\lag\bar{q}g\si Gq\rag}

\def\si{\sigma}

\def\lb{\label}

\newcommand{\rag}{\rangle}
\newcommand{\lag}{\langle}

\journal{Elsevier}

\begin{document}

\begin{frontmatter}

\title{A fresh look into $\overline{m}_{c,b}(\overline{m}_{c,b})$ and precise $f_{D_{(s)},B_{(s)}}$  from  heavy-light QCD spectral sum rules 
$^*$ \corref{cor0}
}
\cortext[cor0]{Some results of this work have been presented at the 16th QCD International Conference (QCD12) , Montpellier, 2-6th july 2012 \cite{SNQCD12}.}
 \author[label2]{Stephan Narison}
\address[label2]{Laboratoire
Particules et Univers de Montpellier, CNRS-IN2P3, 
Case 070, Place Eug\`ene
Bataillon, 34095 - Montpellier, France.}
   \ead{snarison@yahoo.fr}
\begin{abstract}
\nin
Using recent values of the QCD (non-) perturbative parameters given in Table \ref{tab:param} and an estimate of the N3LO QCD perturbative contributions based on the geometric growth of the PT series, we re-use QCD spectral sum rules (QSSR) known to N2LO PT series and including all dimension-six NP condensate contributions in the full QCD theory,   for improving the existing estimates of $\overline{m}_{c,b}$ and $f_{D_{(s)}, B_{(s)}}$ from  the open charm and beauty systems.  We especially study the effects of the subtraction point on ``different QSSR data" and use (for the first time) the Renormalization Group Invariant (RGI) scale independent quark masses in the analysis. The estimates [rigourous model-independent upper bounds within the SVZ framework] reported in Table~\ref{tab:result}: $f_D/f_\pi=1.56(5)[\leq 1.68(1)]$,  $f_B/f_\pi=1.58(5)[\leq 1.80(3)]$ and $f_{D_s}/f_K= 1.58(4) [\leq 1.63(1)]$, 
$f_{B_s}/f_K=1.50(3)[\leq 1.61(3.5)]$, which improve previous QSSR estimates, 
are in perfect agreement (in values and precisions) with some of the experimental data on $f_{D,D_s}$ and  on recent  lattice simulations within dynamical quarks. 
These remarkable agreements confirm both the success of the QSSR semi-approximate approach based on the OPE in terms of the quark and gluon condensates and of the Minimal Duality Ansatz (MDA) for parametrizing the hadronic spectral function which we have tested from the complete data of the $J/\psi$ and $\Upsilon$ systems.  The values of the running quark masses $\overline{m}_c({m_c})=1286(66) $ MeV and  $\overline{m}_b({m_b})= 4236(69)$ MeV from  $M_{D,B}$ are in good agreement though less accurate than the ones from recent $J/\psi$ and $\Upsilon$ sum rules. 
\end{abstract}
\end{frontmatter}
\section{Introduction and a short historical review}
The (pseudo)scalar meson decay constants $f_P$ are of prime interests for understanding the realizations of chiral symmetry in QCD. 
In addition to the well-known values of $f_\pi=(130.4(2)$ MeV and $f_K=156.1(9)$ MeV \cite{ROSNER} which control  the light flavour chiral symmetries, it is also desirable to extract the ones of the heavy-light charm and bottom quark systems with high-accuracy. These decay constants are normalized through the matrix element:
\beq
\la 0|J^P_{\bar qQ}(x)|P\ra= f_P M_P^2~,
\label{eq:fp}
\eeq
where: 
\beq
J^P_{\bar qQ}(x)\equiv (m_q+M_Q)\bar q(i\gamma_5)Q~,
\label{eq:current}
\eeq
is the local heavy-light pseudoscalar current;  $q\equiv d,s;~Q\equiv c,b;~P\equiv D_{(s)}, B_{(s)}$and where $f_P$  is related to  the leptonic width:
\beq
\Gamma (P^+\to l^+\nu_l)={G^2_F\over 8\pi}|V_{Qq}|^2f^2_Pm_l^2M_P\ga 1-{m_l^2\over M_P^2}\dr^2~,
\eeq
where $m_l$ is the lepton mass and $|V_{Qq}|$ the CKM mixing angle. 
Besides some earlier attempts based on non-relativistic potential models to extract these quantities (which are however not applicable for the heavy-light systems), the first bounds on $f_D$ and $f_B$ from QCD spectral sum rules (QSSR) \cite{SVZ}\,\footnote{For reviews, see e.g: \cite{RRY,SNB1,SNB2,SNB3}.} were derived by NSV2Z \cite{NSV2Z}, which have been improved four years later in \cite{SNZ, BROAD,GENERALIS}. Since then, but long time before the lattice results, different QSSR papers have been published in the literature for estimating $f_{D,B}$\,\footnote{For reviews and more complete references, see e.g:\cite{SNB1,SNB2}.}. These results look, at first sight, in disagreement among each others and some of them, claimed the observation of the scaling $f_P\sim 1/\sqrt{M_P}$ expected in the large $M_P$ limit \cite{HQET}.
These different papers have been scrutinized  in \cite{SNFB,SNB2}, where Narison found that the apparent discrepancies between the different results can be solved if one applies carefully the stability  criteria (also called sum rule window) of the results versus the external QSSR Laplace/Moments sum rules variables and continuum threshold $t_c$. In this way, and for given values of $m_{c,b}$, he obtained the values:
\beq
f_D\simeq (1.31\pm 0.12)f_\pi~,~~~~~ f_B\simeq (1.6\pm 0.1)f_\pi~,
\label{eq:fdb}
\eeq
which are independent of the forms of the sum rules used. However, the result has been quite surprising as it indicates a large violation of the heavy quark symmetry scaling predictions,where $1/M_Q$ corrections have been estimated in \cite{SNFB4}.  This ``unexpected result" has been confirmed few years later by lattice calculations \cite{MARTI}. 
Since then, some progresses have been done for improving the QCD expression of the 2-point correlator. It starts from a confirmation of the SVZ original expression of the LO perturbative and non-perturbative contributions. Then, Broadhurst and Generalis \cite{BROAD,BROAD1} have provided the complete PT $\alpha_s$ NLO including light quark mass corrections. It has been completed by the PT $\alpha_s^2$ N2LO corrections of Chetyrkin and Steinhauser \cite{CHET} in the case of one heavy and one massless quarks.
This result has been completed by the inclusion of the NP contributions up to dimension-six \cite{GENERALIS} and of the light quark mass corrections to LO by \cite{GENERALIS,JAMIN}. All of these previous QCD expressions have been given in terms of the on-shell quark mass.
In \cite{SNFB2}, Narison has used (for the first time) the running $c,b$ quark masses in the QSSR analysis, by using its known relation with the on-shell mass known at present to NLO \cite{TAR,COQUE,SNPOLE}, N2LO \cite{BROAD,BROAD1} and N3LO \cite{CHET2} where it has been noticed that the QSSR PT expressions converge faster. It has also been noticed that the values of $f_{D,B}$ are very sensitive to the value of $m_{c,b}$ motivating him to extract $m_{c,b}$ (for the first time) from the known values of $M_D$ and $M_B$. 
Recent analysis, including the $\alpha_s^2$ corrections have been presented in the literature, in the full theory where the running $\overline{MS}$ mass has been used \cite{SNFB3,JAMIN3,KHOJ} and in HQET~\cite{PENIN} where the radiative corrections are large due to the uses of the on-shell mass~\footnote{We plan to analyze the HQET sum rules \cite{SNFB4,NEUBERT} in a separate publication.}. \\
In the following, we shall present analysis based on the full QCD theory where we use as inputs the most recent values of the (non-)perturbative QCD parameters given in Table \ref{tab:param}. We assume the  geometric growth of the PT series \cite{NZ} as a dual to the effect of a $1/q^2$ term \cite{CNZ,ZAK} for an estimate of the N3LO perturbative contributions. We shall also study systematically the effect of the substraction points on each ``QSSR data" and use (for the first time) in the analysis, the Renormalization Group Invariant (RGI) $s,c,b$ quark masses introduced by \cite{FNR} and which are scale and (massless) scheme independent.
\section{QCD spectral  sum rules (QSSR)}
\subsection*{\b The Laplace  sum rules (LSR)}
\nin
We shall be concerned with the two-point correlator :
\beq
\psi^{P}_{\bar qQ}(q^2)=i\int d^4x ~e^{iq.x}\lag 0
|TJ^P_{\bar qQ}(x)J^P_{\bar qQ}(0)^\dagger
|0\rag~,
\lb{2po}
\eeq
where $J_{\bar qQ}(x)$ is the local current defined in Eq. (\ref{eq:current}). 
The associated Laplace sum rules (LSR)  ${\cal L}_{\bar qQ}(\tau)$ and
its ratio ${\cal R}_{\bar qQ}(\tau)$ read\,\cite{SVZ}\,\footnote{Radiative corrections to the exponential sum rules have been first derived in \cite{SNRAF}, where it has been noticed that the PT series has the property of an Inverse Laplace transform.}:
\beq
{\cal L}_{\bar qQ}(\tau,\mu)=\int_{(m_q+M_Q)^2}^{t_c}dt~e^{-t\tau}\frac{1}{\pi} \mbox{Im}\psi^P_{\bar qQ}(t,\mu)~,
\label{eq:lsr}
\eeq
\beq\label{eq:ratiolsr}
{\cal R}_{\bar qQ} (\tau,\mu) = \frac{\int_{(m_q+M_Q)^2}^{t_c} dt~t~ e^{-t\tau}\frac{1}{\pi}\mbox{Im}\psi^P_{\bar qQ}(t,\mu)}
{\int_{(m_q+M_Q)^2}^{t_c} dt~ e^{-t\tau} \frac{1}{\pi} \mbox{Im}\psi_{\bar qQ}(t,\mu)}~,
\eeq
where $\mu$ is the subtraction point which appears in the approximate QCD series when radiative corrections are included. 
 The ratio of sum  rules ${\cal R}_{\bar qQ} (\tau,\mu)$ is useful, as it is equal to the
resonance mass squared, in
  the Minimal Duality Ansatz (MDA) parametrization of the spectral function:
\beq
\frac{1}{\pi}\mbox{ Im}\psi^P_{\bar qQ}(t)\simeq f^2_PM_P^4\delta(t-M^2_P)
  \ + \
  ``\mbox{QCD cont.}" \theta (t-t_c),
\label{eq:duality}
\eeq
where $f_P$ is the decay constant defined in Eq. (\ref{eq:fp}) and the higher states contributions are smeared by the ``QCD continuum" coming from the discontinuity of the QCD diagrams and starting from a constant threshold $t_c$. 
\subsection*{\b The $Q^2=0$ moment  sum rules (MSR)}
\nin
We shall also use for the $B$-meson, the moments obtained after deriving $n+1$-times the two-point function and evaluated at $Q^2=0$ \cite{SVZ}, where an expansion in terms of the on-shell mass $M_b$ can be used. They read:
\beq
{\cal M}^{(n)}_{\bar qb}(\mu)=\int_{(m_q+M_b)^2}^{t_c}{dt\over t^{n+2}}~\frac{1}{\pi} \mbox{Im}\psi^P_{\bar qb}(t,\mu)~,
\label{eq:mom}
\eeq
and the associated ratio:
\beq
{\cal R}^{(n)}_{\bar qb}(\mu)={\int_{(m_q+M_b)^2}^{t_c}{dt\over t^{n+2}}~\frac{1}{\pi} \mbox{Im}\psi^P_{\bar qb}(t,\mu)\over 
\int_{(m_q+M_b)^2}^{t_c}{dt\over t^{n+3}}~\frac{1}{\pi} \mbox{Im}\psi^P_{\bar qb}(t,\mu)}~.
\label{eq:momratio}
\eeq
\begin{figure}[hbt] 
\begin{center}
\centerline {\hspace*{-7.5cm} a) }\vspace{-0.6cm}
{\includegraphics[width=7cm  ]{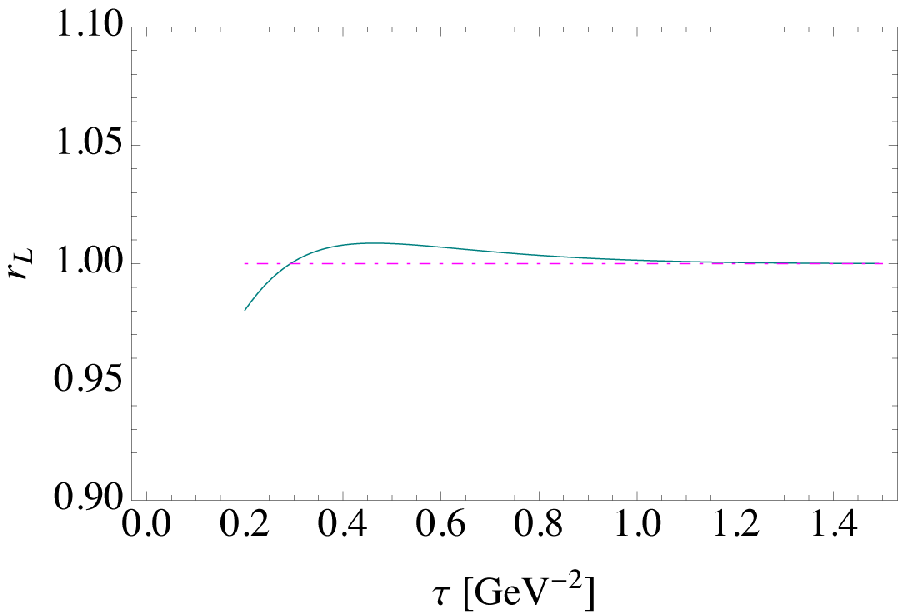}}
\centerline {\hspace*{-7.5cm} b) }\vspace{-0.3cm}
{\includegraphics[width=7cm  ]{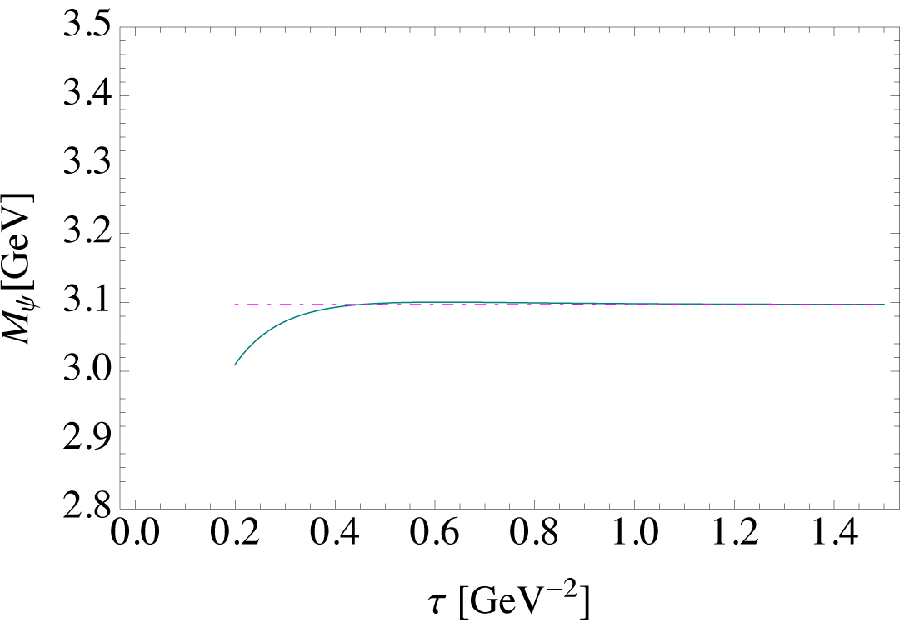}}
\caption{
\scriptsize 
{\bf a)} $\tau$-behaviour of the ratio of  ${\cal L}_{\bar cc}^{exp}/ {\cal L}_{\bar cc}^{dual}$ for $\sqrt{t_c}= M_{\psi(2S)}$-0.15 GeV. The red dashed curve corresponds to the strict equality for all values of $\tau$.; {\bf b)} the same as a) but for $M_\psi=\sqrt{{\cal R}_{\bar cc}}$. 
}
\label{fig:cduality}
\end{center}
\end{figure} 
\nin
\begin{figure}[hbt] 
\begin{center}
\centerline {\hspace*{-7.5cm} a) }\vspace{-0.6cm}
{\includegraphics[width=7cm  ]{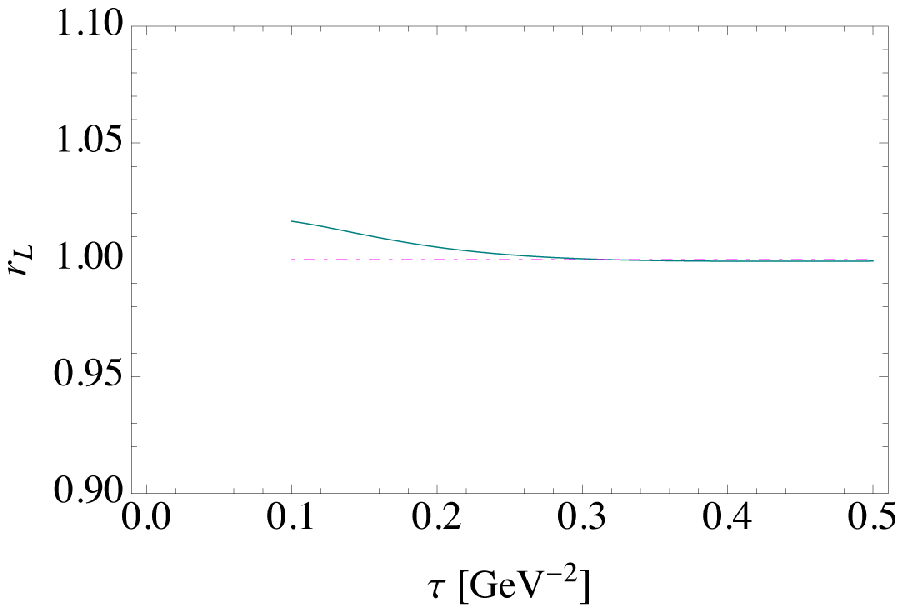}}
\centerline {\hspace*{-7.5cm} b) }\vspace{-0.3cm}
{\includegraphics[width=7cm  ]{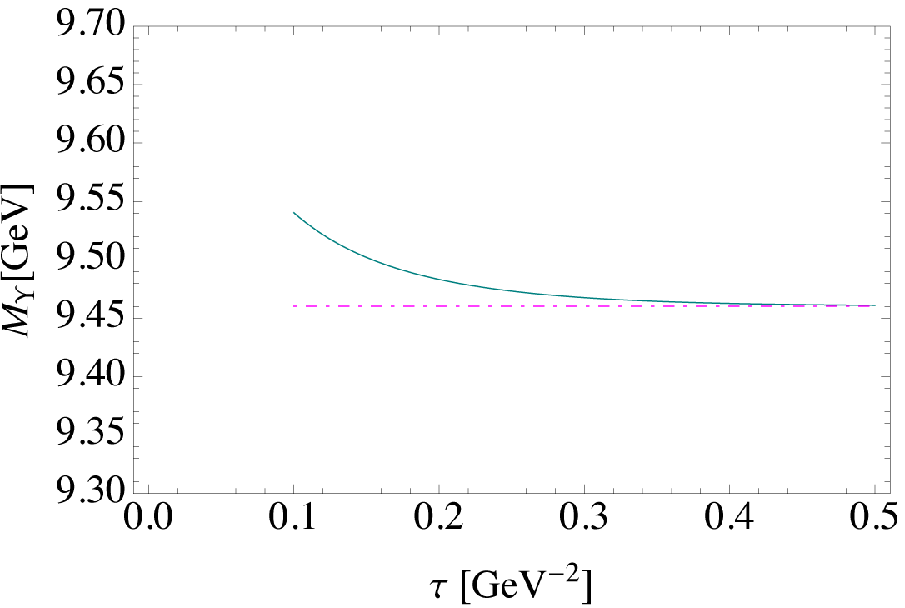}}
\caption{
\scriptsize 
The same as in Fig. \ref{fig:cduality} but for the $b$-quark and for $\sqrt{t_c}= M_{\Upsilon(2S)}$-0.15 GeV.
}
\label{fig:bduality} 
\end{center}
\end{figure} 
\nin
\subsection*{\b Test of the Minimal Duality Ansatz (MDA) from $J/\psi$ and $\Upsilon$}\label{sec:duality}
\begin{figure}[hbt] 
\begin{center}
\centerline {\hspace*{-7.5cm} a) }\vspace{-0.6cm}
{\includegraphics[width=7cm  ]{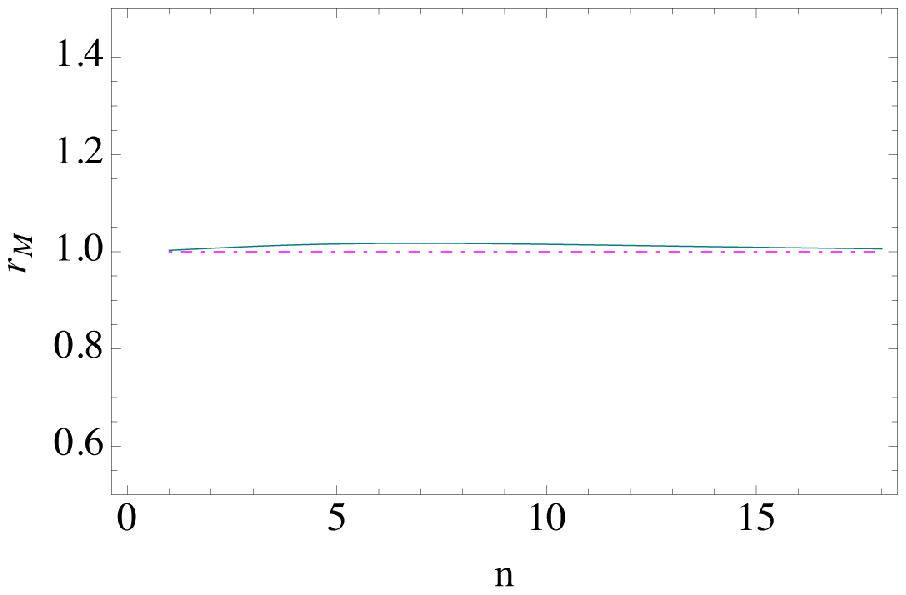}}
\centerline {\hspace*{-7.5cm} b) }\vspace{-0.3cm}
{\includegraphics[width=7cm  ]{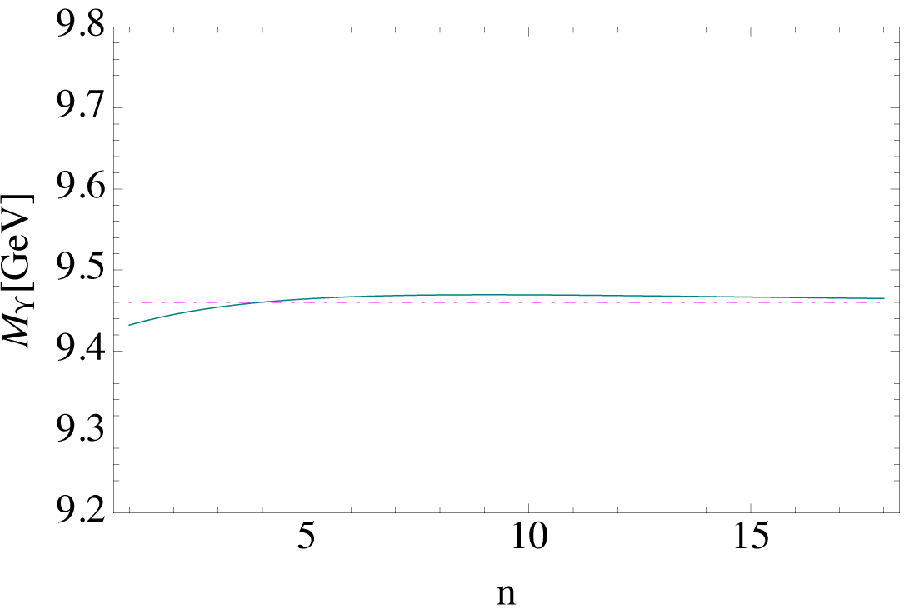}}
\caption{
\scriptsize 
The same as in Fig. \ref{fig:bduality} but for the $Q^2=0$ moment of the $b$-quark versus the number of derivatives $n$.
}
\label{fig:bmom_duality} 
\end{center}
\end{figure}

\nin
 We have checked explicitly in \cite{SNB2} that the MDA presented in Eq. (\ref{eq:duality}), when applied to the $\rho$-meson reproduces within 15\% accuracy the ratio ${\cal R}_{\bar dd}$ measured from the total cross-section $e^+e^-\to {\rm I=1 ~hadrons}$ data (Fig. 5.6 of \cite{SNB2}). In the case of charmonium, we have also compared $M_\psi^2$ from ${\cal R}^{(n)}_{\bar cc}$ with the one from complete data and find a remarkable agreement for higher $n\geq 4$ values (Fig. 9.1 of \cite{SNB2}), indicating that for heavy quark systems the r\^ole of the QCD continuum will be smaller than in the case of light quarks and the exponential weight or high number of derivatives suppresses efficiently the QCD continuum contribution but enhances the one of the lowest ground state in the spectral integral. 
 We redo the test done for charmonium in Fig. 9.1 of \cite{SNB2} and analyze the bottomium channel for the LSR and MSR. We show in Fig. (\ref{fig:cduality}a) the $\tau$-behaviour of the ratio of ${\cal L}^{exp}_{\bar cc}$ normalized to  ${\cal L}^{dual}_{\bar cc}$ where we have used the simplest QCD continuum 
 expression for massless quarks to order $\alpha_s^3$ from the threshold $t_c$\,\footnote{We have checked that the spectral function including complete mass corrections give the same results.}:
 \beq
 {\rm QCD~ cont.}=1+as+1.5as^2-12.07as^3.
 \eeq
 We show in Fig. (\ref{fig:cduality}b) the $\tau$-behaviour of $M_\psi$, where the continuous (oliva) curve corresponds to $\sqrt{t_c}\simeq M_{\psi(2S)}- 0.15$ GeV. We show a similar analysis for the bottomium sum rules in Fig. (\ref{fig:bduality}) for the LSR and in Fig. (\ref{fig:bmom_duality}) for the MSR where we have taken $\sqrt{t_c}\simeq M_{\Upsilon(2S)}- 0.15$ GeV. One can see that the MDA, with a value of $\sqrt{t_c}$ around the value of the 1st radial excitation mass, describes quite well the complete data in the region of $\tau$ and $n$ where the corresponding sum rules present $\tau$ or $n$ stability\,\cite{SNH10}:
 \bea
  \tau^\psi&\simeq& (1.3\sim 1.4)~{\rm GeV}^{-2},\nnb\\
  ~ \tau^\Upsilon&\simeq& (0.2\sim 0.4)~{\rm GeV}^{-2},~~~~~~~n^\Upsilon\simeq (5\sim 7)~, 
  \eea
  as we shall see later on. 
 This good description of the data by the MDA shows the efficient r\^ole of the exponential weight or high number of derivatives for suppressing the higher mass states and QCD continuum contribution in the analysis. This nice feature prevents the introduction of some more involved models bringing new parameters in the analysis where some of them cannot be understood from QCD 1st  principles. 
 Moreover, MDA  has been also used in \cite{PERIS} (called Minimal Hadronic Ansatz in this paper) in the context of large $N_c$ QCD, where the restriction of an infinite set of large $N_c$ narrow states to a Minimal Hadronic Ansatz which is needed to satisfy the leading short- and long-distance behaviours o the relevant Green's functions, provides a very good approximation to the observables one compute. 
\nin
\begin{figure}[hbt] 
\begin{center}
{\includegraphics[height=7cm  ]{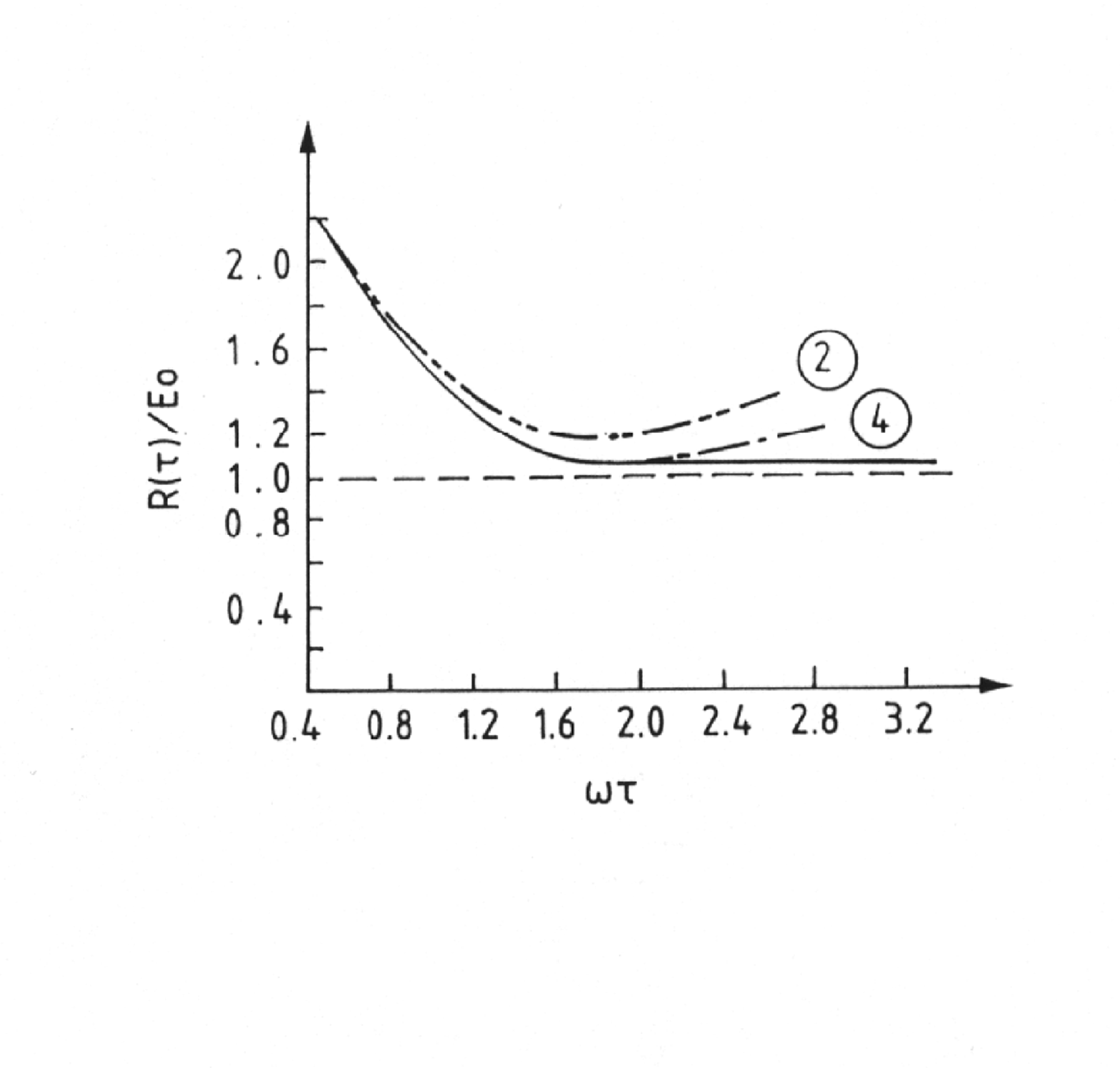}}
\vspace*{-2cm}
\caption{
\scriptsize 
{\bf a)} $\tau$-behaviour of ${\cal R}(\tau)$ normalized to the ground state energy $E_0$ for the harmonic oscillator. 2 and 4 indicate the number of terms in the approximate series.}
\label{fig:oscillo} 
\end{center}
\end{figure} 
\nin
\subsection*{\b Optimal results from stability criteria}
\nin
Using the theoretical expressions of ${\cal L}_{\bar dQ}^{th}$ or ${\cal M}_{\bar db}^{(n)th}$, and parametrizing its experimental side ${\cal L}_{\bar dQ}^{exp}$ or ${\cal M}_{\bar db}^{(n)exp}$ by the MDA in  Eq. (\ref{eq:duality}), one can extract the decay constant $f_P$ and the RGI quark mass $\hat m_Q$.
In principle the equality  ${\cal L}_{\bar dQ}^{th}={\cal L}_{\bar dQ}^{exp}$ should be satisfied for any values of the external (unphysical) set of variables $(\tau, t_c)$, if one knows exactly ${\cal L}_{\bar dQ}^{th}$ and ${\cal L}_{\bar dQ}^{exp}$. Unlike the harmonic oscillator, this is not the case. Using  the ratio of moments ${\cal R}_{\bar dQ}$ for the harmonic oscillator as a function of the imaginary time variable $\tau$, where one knows the exact and approximate results, one can find \cite{BELL} that the exact energy $E_0$ of the ground state can be approached from above by the approximate series
(see Fig. \ref{fig:oscillo}).  At the minimum or inflexion point (stability) of the curves, one has a ground state dominance. For small time (large $Q^2$), all level contributes, while for large time (small $Q^2$) the series breakdown.
We shall apply this stability criterion inspired from quantum mechanics in our analysis. \\
In principle, the continuum threshold $\sqrt{t_c}$ in  Eq. (\ref{eq:duality}) is a free parameter, though one expects its value to be around the mass of the 1st radial excitation because the QCD spectral function is supposed to smear all the higher state contributions in the spectral integral as explicitly shown previously in Section \ref{sec:duality}. In order to avoid the model-dependence on the results, Refs. \cite{SNFB,SNFB2,SNFB3,SNFB4,SNB1,SNB2} have considered the conservative range of $t_c$-values where one starts to have $\tau$- or $n$-stability until which one reaches a $t_c$-stability where the contribution of the lowest ground state to the spectral integral completely dominates. For the $D$ and $B$ mesons, this range is \cite{SNFB,SNFB2,SNFB3,SNFB4,SNB1,SNB2}:
\beq
t_c^D\simeq (5.5 \to 9.5)~{\rm GeV}^2,~~~~~t_c^B\simeq (33\to 45) ~{\rm GeV}^2.
\eeq
\section{The QCD input parameters}
\nin
The QCD parameters which shall appear in the following analysis will be the strange, charm and bottom quark masses $m_{s,c,b}$ (we shall neglect  the light quark masses $q\equiv u,d$),
the light quark condensate $\qq$,  the gluon condensates $ \lag
g^2G^2\rag\equiv \la g^2G^a_{\mu\nu}G_a^{\mu\nu}\ra$ and $\la g^3G^3\ra\equiv \la g^3f_{abc}G^a_{\mu\nu}G^b_{\nu\rho}G^c_{\rho\mu}\ra$, the mixed condensate $\mix\equiv {\la\bar qg\sigma^{\mu\nu} (\lambda_a/2) G^a_{\mu\nu}q\ra}=M_0^2\la \bar qq\ra$ and the four-quark 
 condensate $\rho\la\bar qq\ra^2$, where
 $\rho\simeq 2$ indicates the deviation from the four-quark vacuum 
saturation. Their values are given in Table {\ref{tab:param}} and we shall work with the running
light quark parameters known to order $\alpha_s^3$ \cite{SNB1,SNB2,RUNDEC}. They read:
\bea
{\bar m}_{q,Q}(\tau)&=&
{\hat m}_{q,Q}  \ga-\beta_1a_s\dr^{-2/{
\beta_1}}\times C(a_s)
\nnb\\
{\la\bar qq\ra}(\tau)&=&-{\hat \mu_q^3  \ga-\beta_1a_s\dr^{2/{
\beta_1}}}/C(a_s)
\nnb\\
{\la\bar qg\sigma Gq\ra}(\tau)&=&-{M_0^2{\hat \mu_q^3} \ga-\beta_1a_s\dr^{1/{3\beta_1}}}/C(a_s)~,
\eea
where $\beta_1=-(1/2)(11-2n_f/3)$ is the first coefficient of the $\beta$ function 
for $n_f$ flavours; $a_s\equiv \alpha_s(\tau)/\pi$; ${\hat m}_{q,Q}$ is the RGI quark mass, $\hat\mu_q$ is spontaneous RGI light quark condensate \cite{FNR}. The QCD correction factor $C(a_s)$ in the previous expressions is numerically:
\bea
C(a_s)&=& 1+0.8951a_s+1.3715a_s^2 +...~~{\rm :}~~ n_f=3~,\nnb\\
&=&1+1.1755a_s+1.5008a_s^2 +...~~{\rm :}~~ n_f=5~,
\eea
which shows a good convergence. 
We shall use:
\beq   
\alpha_s(M_\tau)=0.325(8) \lrar  \alpha_s(M_Z)=0.1192(10)
\label{eq:alphas}
\eeq
from $\tau$-decays \cite{SNTAU,BNP}, which agree perfectly with the world average 2012 \cite{BETHKE,PDG}: 
\beq
\alpha_s(M_Z)=0.1184(7)~. 
\eeq
We shall also use the value of the running strange quark mass obtained in \cite{SNmass}\,\footnote{This value agrees and improves previous sum rules results \cite{SNmass2}.} given in Table~\ref{tab:param}. The value of the running $\la \bar qq\ra$ condensate is deduced from the value of $(\overline{m}_u+\overline{m}_d)(2)=(7.9\pm 0.6)$ MeV obtained in  \cite{SNmass} and the well-known GMOR relation: $(m_u+m_d)\la \bar uu+\bar dd\ra=-m_\pi^2f_\pi^2$. The values of the running $\overline{MS}$ mass $\overline{m}_Q(M_Q)$  recently obtained in Ref. \cite{SNH10} from charmonium and bottomium sum rules, will also be used\,\footnote{These values agree and improve previous sum rules results \cite{SVZ,RRY,SNB1,SNB2,SNHmass,IOFFE}.}. Their average is given in Table~\ref{tab:param}. From which, we deduce the RGI invariant heavy quark masses to order $\alpha_s^2$, in units of MeV:
\beq
\hat m_c=1467(14) ~,~~~~~~~~~~\hat m_b=7292(14) ~.
\label{eq:heavymass}
\eeq
For the light quarks, we shall use the value of the RGI mass and spontaneous mass to order $\alpha_s$ for consistency with the known $\alpha_s$ $m_s$ and $\la \bar qq\ra $ condensate corrections of the two-point correlator. They read, in units of MeV:
\beq
\hat m_s=128(7) ~,~~~~~~~~~~~~~\hat\mu_q=251(6)~.
\label{eq:lightmass}
\eeq
{\scriptsize
\begin{table}[hbt]
\setlength{\tabcolsep}{0.2pc}
 \caption{
QCD input parameters. 
 }
    {\small
\begin{tabular}{lll}
&\\
\hline
Parameters&Values& Ref.    \\
\hline
$\alpha_s(M_\tau)$& $0.325(8)$&\cite{SNTAU,BNP,BETHKE}\\
$\overline{m}_s(2)$&$96.1(4.8)$ MeV&average \cite{SNmass}\\
$\overline{m}_c(m_c)$&$1261(12)$ MeV &average \cite{SNH10}\\
$\overline{m}_b(m_b)$&$4177(11)$ MeV&average \cite{SNH10}\\
${1\over 2}\la \bar uu+\bar dd\ra^{1/3} (2)$&$-(275.7\pm 6.6)$ MeV&\cite{SNB1,SNmass}\\
$\la\bar ss\ra/\la\bar dd\ra$&0.74(3)&\cite{SNB1,SNmass,Hbaryon}\\
$M_0^2$&$(0.8 \pm 0.2)$ GeV$^2$&\cite{JAMI2,HEID,SNhl}\\
$\la\alpha_s G^2\ra$& $(7\pm 1)\times 10^{-2}$ GeV$^4$&
\cite{SNTAU,LNT,SNI,fesr,YNDU,SNHeavy,BELL,SNH10,SNG}\\
$\la g^3  G^3\ra$& $(8.2\pm 1.0)$ GeV$^2\times\la\alpha_s G^2\ra$&
\cite{SNH10}\\
$\rho \la \bar qq\ra^2$&$(4.5\pm 0.3)\times 10^{-4}$ GeV$^6$&\cite{SNTAU,LNT,JAMI2}\\
\hline
\end{tabular}
}
\label{tab:param}
\end{table}
}
\section{QCD expressions of the  sum rules}

\subsection*{\b The  LSR}
\nin
To order $\alpha_s^2$, the QCD theoretical side of the sum rule reads, in terms of the on-shell heavy quark mass $M_Q$ and for $m_d=0$:
\bea\label{eq: lsrlh}
{\cal L}_{\bar dQ}(\tau)
&=& M^2_Q\Bigg{\{}\int_{M^2_Q}^{\infty} 
{dt}~\mbox{e}^{-t\tau}\frac{1}{\pi} \mbox{Im}\psi^P_{\bar qQ}(t)\big{\vert}_{PT}
+ {\la \als G^2\ra\over 12\pi}\mbox{e}^{-z}\nnb\\
&&-\Bigg{\{}\Bigg{[} 1+2a_s \Big{[}1+(1-z)\ga \ln{\nu^2\tau}+{4\over 3}\dr\Big{]}\Bigg{]}~\mbox{e}^{-z}\nnb\\
&&-2a_s\Gamma(0,z)\Bigg{\}}\ga \overline{m}_Q\over M_Q\dr^2\overline{m}_Q\la \bar dd\ra~, \nnb\\
&&-\tau ~\mbox{e}^{-z}\Bigg{\{}{z\over 2}\ga 1-\frac{z}{2}\dr
M_QM_0^2\la\bar dd\ra\nnb\\ 
&&+\ga 2-\frac{z}{2}-\frac{z^2}{6}\dr{\la\bar dj d\ra\over 6}\nnb\\
&&-\ga 1+z-7z^2+{5\over 3}z^3\dr {\la g^3 G^3\ra\over 2880\pi^2}\nnb\\
&&+\Bigg{[} 5\tilde L (12-3z-z^2)z-9+11z+{41\over 2}z^2\nnb\\
&&+{5\over 2}z^3\Bigg{]}
{ \la j^2\ra\over 2160\pi^2}\Bigg{\}}~,
\eea
where:
\beq
\mbox{Im}\psi^P_{\bar qQ}(t)\big{\vert}_{PT}=\frac{1}{8\pi^2}\Bigg{[} 3 t(1-x)^2
\ga
1+\frac{4}{3}a_s f(x)\dr+
a_s^2 R_{2s}\Bigg{]}
\eeq
 with: $z\equiv M_Q^2\tau$; $x\equiv M^2_Q/t$; $a_s\equiv \alpha_s/\pi$; $\tilde L\equiv \ln {(\mu M_Q\tau)}+\gamma_E:  \gamma_E=0.577215...$; $\mu$ is an arbitrary subtraction point; $R_{2s}$ is the $\alpha_s^2$-term obtained semi -analytically in \cite{CHET} and is available as  a Mathematica package program Rvs.m. Neglecting $m_d$, the PT NLO terms read \cite{BROAD}:
\bea
f(x)&=&\frac{9}{4}+2\rm{Li}_2(x)+\log x \log (1-x)\nnb\\
&&-\frac{3}{2}\log 
(1/x-1) -\log (1-x)+ \nnb\\ 
&& x\log (1/x-1)-(x/(1-x))\log x~.
\eea
The contribution up to the $d=4$ gluon condensate and up to $d=6$ quark condensates have been 
obtained originally by NSV2Z \cite{NSV2Z}. The contribution of the $d=6$ $\la g^3f_{abc}G^3\ra$ and $ \la j^2\ra$ gluon condensates have been deduced from the expressions given by \cite{GENERALIS} (Eqs. II.4.28 and Table II.8) where:
\bea
\la\bar dj d\ra&\equiv&\la \bar d g\gamma_\mu D^\mu G_{\mu\nu}{\lambda_a\over 2} d\ra=
g^2\la \bar d \gamma_\mu {\lambda_a\over 2} d\sum_q \bar q \gamma_\mu {\lambda_a\over 2} q\ra\nnb\\
&\simeq&-{16\over 9} (\pi\alpha_s)~\rho\la \bar dd\ra^2,\nnb\\
 \la j^2\ra&\equiv& g^2 \la (D_\mu G^a_{\nu\mu})^2\ra
=g^4\la\ga \sum_q\bar q\gamma_\nu {\lambda^a\over 2}q\dr^2\hspace*{-0.1cm}\ra \nnb\\
&\simeq& -{64\over 3} (\pi\alpha_s)^2 \rho\la\bar dd\ra^2,
\eea
after the use of the equation of motion. $\rho \simeq (2\pm 0.2)$ measures the deviation from the vacuum saturation estimate of the $d=6$ quark condensates \cite{SNTAU,LNT,JAMI2}. \\
The $\alpha_s$ correction to $\la \bar dd\ra$, in the $\overline{MS}$-scheme, comes  from \cite{JAMIN3}, where the running heavy quark mass  $\overline{m}_Q$ enters into this expression.
Using the known relation between the running $\bar{m}_Q(\mu)$ and on-shell mass
$M_Q$ in the $\overline{MS}$-scheme to order $\alpha_s^2$ \cite{TAR,COQUE,SNPOLE,BROAD2,CHET2}:
\bea
M_Q &=& \overline{m}_Q(\mu)\Big{[}
1+{4\over 3} a_s+ (16.2163 -1.0414 n_l)a_s^2\nnb\\
&&+\ln{\ga\mu\over M_Q\dr^2} \ga a_s+(8.8472 -0.3611 n_l) a_s^2\dr\nnb\\
&&+\ln^2{\ga\mu\over M_Q\dr^2} \ga 1.7917 -0.0833 n_l\dr a_s^2+...\Big{]}~,
\label{eq:pole}
\eea
for $n_l$ light flavours, one can express all terms of the previous sum rules with the running mass $\overline{m}_Q(\mu)$.
It is clear that, for some non-perturbative terms which are known to leading order
of perturbation theory, one can use either the running or the pole 
mass. However, we shall see that this distinction does not affect, in a visible way, the present result, within the accuracy of our estimate, as the non-perturbative contributions are relatively small though vital in the analysis. 
\subsection*{\b The MSR}
\nin
The moments read for $m_d=0$:
\bea
{\cal M}^{(n)}_{\bar db}&=&\int_{M_b^2}^{t_c}{dt\over t^{n+2}}~\frac{1}{\pi} \mbox{Im}\psi^B_{\bar db}(t)\big{\vert}_{PT}+\nnb\\
&&{1\over \ga M_b^2\dr^{n+1}}\Bigg{\{}-M_b\la\bar dd\ra+{\la \als G^2\ra\over 12\pi}\nnb\\
&&+(n+1)(n+2) {1\over 4M_b} M_0^2\la\bar dd\ra\nnb\\
&&-(n+1)(n+2)(n+9){1\over M_b^2}{\la \bar d j d\ra \over 36}\nnb\\
&&- (n+3) (5 n^2+9 n+1){1\over 3M_b^2}{\la g^3G^3\ra \over  2880\pi^2}\nnb\\
&&-\Bigg{\{}{1\over 3}(20 n^3 + 186 n^2 + 337 n + 117)+\nnb\\
&&-5(n+2)\Big{[} S_4(n^2+7n+12)+\nnb\\
&& \hspace*{1.5cm}3S_3(n+3)-12S_2-\nnb\\
&&(n^2+10n+9)\ln{\ga M_b\over\mu\dr}\Big{]}
\Bigg{\}}{1\over M_b^2}{ \la j^2\ra\over 2160\pi^2}\Bigg{\}}~,
\label{eq:mom2}
\eea
 where:
 \beq
 S_p\equiv \sum_{i=0}^n{{1\over i+p}}~.
\eeq
\begin{figure}[hbt] 
\begin{center}
\centerline {\hspace*{-7.5cm} a) }\vspace{-0.6cm}
{\includegraphics[width=7cm  ]{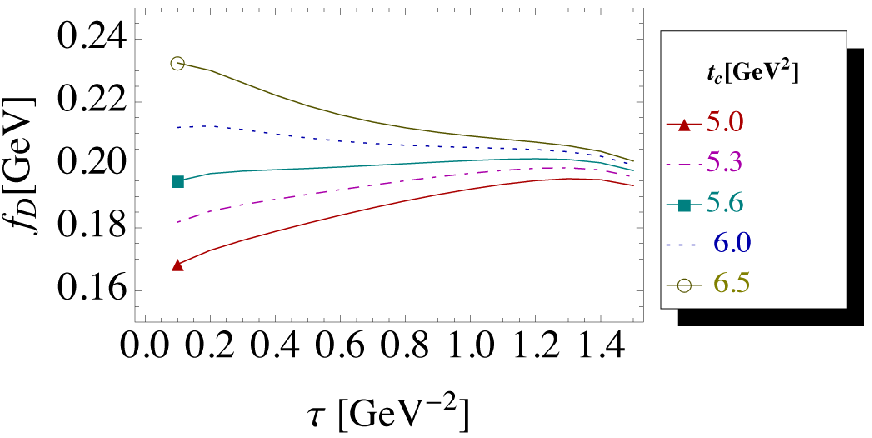}}
\centerline {\hspace*{-7.5cm} b) }\vspace{-0.3cm}
{\includegraphics[width=7cm  ]{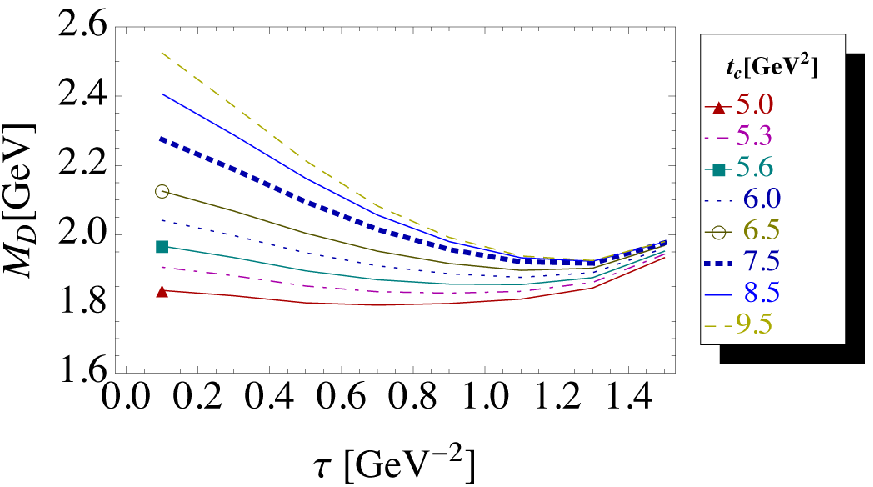}}
\caption{
\scriptsize 
{\bf a)} $\tau$-behaviour of $f_D$ from  ${\cal L}_{\bar dc} $ for different values of $t_c$, for a given value of the subtraction point $\mu=\tau^{-1/2}$ and for $\hat m_c=1467$ MeV as given in Eq. (\ref{eq:heavymass}); {\bf b)} the same as a) but for $M_D$ from ${\cal R}_{\bar dc}$. }
\label{fig:fdtau} 
\end{center}
\end{figure} 
\nin
\section{Estimates of $f_P$ and $\hat m_Q$ at $\mu=\tau^{-1/2}$ from LSR }
After inspection, one finds that $f_P$ and the RGI mass $\hat m_Q$ can only be simultaneously determined from
${\cal L}_{\bar dQ} (\tau,\mu)$ and  ${\cal R}_{\bar dQ} (\tau,\mu)$ evaluated  at $\mu=\tau^{-1/2}$. For other values of $\mu$, only ${\cal L}_{\bar dc} (\tau,\mu)$ present $\tau$ stability at reasonable values of $\tau\leq 1.2$ GeV$^{-2}$, which is not ${\cal R}_{\bar dc} (\tau,\mu)$.
This particular  value of $\mu=\tau^{-1/2}$ is also interesting because the subtraction scale moves with the sum rule variable $\tau$ in the analysis. 
\subsection*{\b Analysis of the $\tau$-and $t_c$-stabilities of ${\cal L}_{\bar dc} $ and ${\cal R}_{\bar dc}$ }
\nin
Using the central values of the QCD input parameters in Table \ref{tab:param} and in Eqs. (\ref{eq:alphas}), (\ref{eq:heavymass}) and  (\ref{eq:lightmass}), one can show in Fig. \ref{fig:fdtau} the influences of
$\tau$ and $t_c$ on the value of $f_D$ and $M_D$  for a given value of the subtraction point $\mu=\tau^{-1/2}$,
where,  the $\tau$-stability for $f_D$ is reached for:
\beq
\tau^D_0\simeq (0.8\sim 1.2)~{\rm GeV}^{-2}~,~~~~t_c^D\simeq (5.3\to 6.5) ~{\rm GeV}^2~
\label{eq:tctau}
\eeq
When extracting  the RGI mass $\hat m_c$ from ${\cal R}_{\bar dc}$ by requiring that it reproduces the experimental mass squared $M_D^2$, one can notice in Fig. (\ref{fig:fdtau}) that, unlike $f_D$, $M_D$ present $\tau$-stability for larger range of $t_c$-values:
\beq
t_c^D\simeq  (5.3 \to 9.5) ~{\rm GeV}^{2}~.
\label{eq:tctaub}
\eeq 
The existence of $\tau$-stability at values of $t_c$ below 5.3 GeV$^2$ depends on the heavy quark mass value and disappears when we require the sum rule to reproduce the value of $M_D$, such that we shall not consider a such region. The values of $t_c\simeq (6.5\sim 9.5)$ GeV$^2$ given in Eqs. (\ref{eq:tctau}) and (\ref{eq:tctaub}) correspond the beginning of $t_c$ stability, where at the extremal values $\tau\simeq (1.2\sim 1.3)$ GeV$^{-2}$, optimal results for $f_D,~M_D$ can be extracted (principle of minimal sensitivity on external variable) and where there is a balance between the continuum (left) and non-perturbative (right) contributions.  (see also similar cases of the harmonic oscillator in Fig. \ref{fig:oscillo} and of the Laplace sum rules for charmonium and bottomium in \cite{SNH10,BELL}). Like in earlier versions of this work \cite{SNFB,SNFB4,SNFB2,SNFB3}, we consider this large range of $t_c$-values in the aim to extract the most conservative result from the analysis and to avoid, in the same way, any (ad hoc) external input  for fixing the exact value of $t_c$. This procedure implies a larger error in our result than often quoted in the literature where (to my personal opinion) the systematics have been underestimated. A similar procedure will be done in the following and for the $B$-meson channel. 
\subsection*{\b Analysis of the convergence of the QCD series }
\begin{figure}[hbt] 
\begin{center}
\centerline {\hspace*{-7.5cm} a) }\vspace{-0.6cm}
{\includegraphics[width=7cm  ]{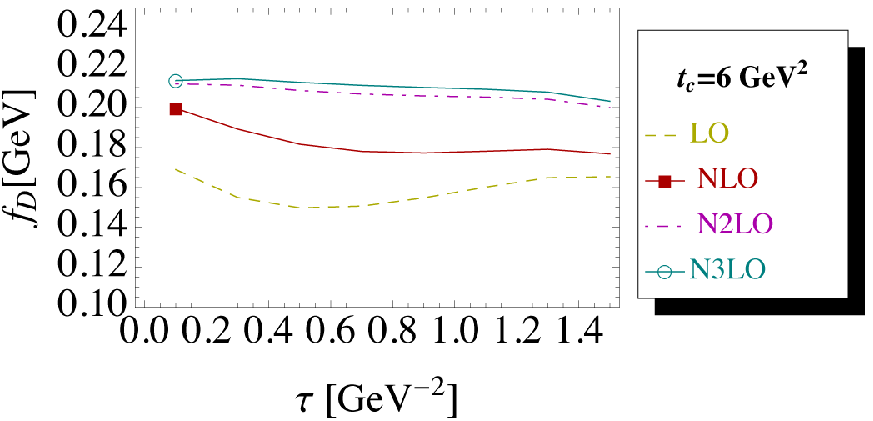}}
\centerline {\hspace*{-7.5cm} b) }\vspace{-0.3cm}
{\includegraphics[width=7cm  ]{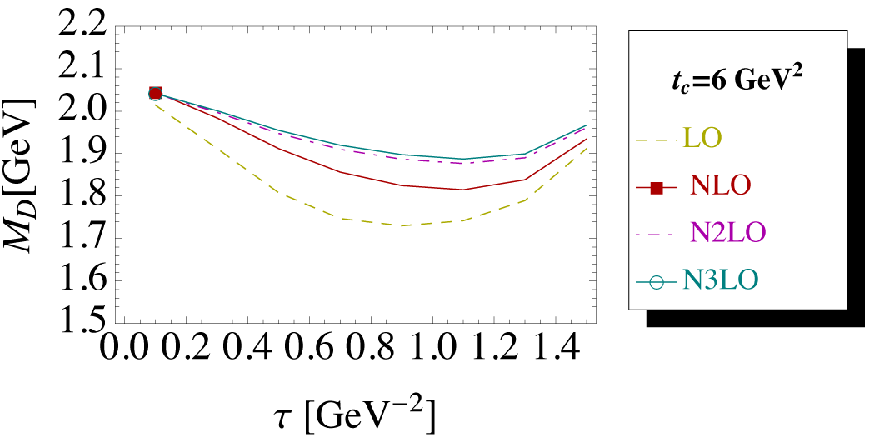}}
\caption{
\scriptsize 
$\tau$-behaviour of $f_D$ from LSR for $t^D_c$= 6 GeV$^2$, for $\hat m_c=1467$ MeV, for a given value of the subtraction point $\mu=\tau^{-1/2}$ GeV and for different truncations of the QCD PT series, where the estimated N3LO contribution is small indicating a good convergence of the series; {\bf b)} the same as a) but for $M_B$.
}
\label{fig:nlo} 
\end{center}
\end{figure}

\nin
We study the convergence of the QCD series in the case of the charm quark at a such low value of the subtraction point $\mu=\tau^{-1/2}$ and taking $t_c=6$ GeV$^2$ . We work in the $\overline{MS}$-scheme as we know from previous works \cite{SNFB2} that the 
PT series converge better than using the on-shell subtraction. In so doing, we estimate the $\alpha_s^3$ N3LO contribution using a geometric PT series as advocated in \cite{NZ} which is dual to the effect of the $1/q^2$ term when large order PT series are resummed. We show the $\tau$-behaviour of $f_D$ in Fig. (\ref{fig:nlo}). One can notice that, all corrections act in a positive way. The prediction increases by about 17\% from LO to NLO and another 14\% from NLO to N2LO but remains unaffected by the inclusion of the N3LO contribution estimated above. These features indicate that the PT series converge quite well at this low scale, while the size of each PT corrections are reasonably small  and will be even smaller for higher values of the subtraction point $\mu$ and  for the  $B$-meson.   Therefore, a confirmation of this N3LO estimate requires an explicit evaluation of this contribution. \\
As far as the non-perturbative contributions are concerned, their effects are relatively small. 
\nin
\subsection*{\b QCD and systematic error estimates }
\nin
Using the previous QCD input parameters and their corresponding errors, we deduce the different errors on $f_P$ and $\hat m_Q$ given in Table \ref{tab:tau}, where the optimal results have been taken at the $\tau$- and $t_c$-stability regions mentioned in the previous subsection:
\beq
\tau^D\simeq (0.8\sim 1.3)~ {\rm GeV}^{-2},~t_c^D\simeq  (5.3 \to 6.5\sim 9.5) ~{\rm GeV}^{2}.
\eeq
As mentioned earlier, we consider a such large range of $t_c$-values in the aim to extract the most conservative result from the analysis. However, this procedure induces a larger error in the analysis than the one quoted in the literature using some other models or using some other criteria. In fact, the range of values of our result  includes most of the predictions given in the literature which are often quoted with smaller errors. Therefore, we expect that, within this procedure, we take properly into account most of the systematics of the sum rule approach. \\
In so doing, we take the central value of $f_D$ in Table \ref{tab:tau} as coming from an arithmetic average of its values from the different $t_c$ given in the legend of Fig. (\ref{fig:fdtau}) inside the range given by Eq. (\ref{eq:tctau}). 
We may have improved the accuracy of our predictions by introducing more model-dependent new parameters for parametrizing the continuum contribution, which we would not do as, in addition to the test performed in Section \ref{sec:duality},  we also want to check  the degree of accuracy of the MDA parametrization for the heavy-light systems by confronting the results obtained in this paper with the some known data on $f_P$ or from lattice simulations. Indeed, such tests  are important as the MDA model  is widely used in the literature for predicting some not yet measured masses of new exotic hadrons like four-quark, molecules \cite{NAVARRA} and hybrid \cite{STEELE} states. However, we do not also try to fix more precisely $t_c$ by e.g. using Finite Energy Sum Rule \cite{fesr} like did the authors in Ref. \cite{NIELSEN} as we want to have more conservative results.
\subsection*{\b Results for $f_D$ and $\hat m_c$ }
\nin
Considering the common range of $t_c$-values for $f_D$ and $M_D$ given in Eq. (\ref{eq:tctau}), we obtain the results
quoted in Table \ref{tab:tau} which come from an arithmetic average of optimal values obtained at different $t_c$ values in Eq. (\ref{eq:tctau})\,\footnote{Using the larger range of $t_c$-values, we would have obtained a slightly different value: $\hat m_c\simeq 1492(102)_{t_c}(82)_{qcd}$~{\rm MeV}, where the errors come respectively from the choice of $t_c$ and QCD parameters given in details in Table \ref{tab:tau}.}:
\bea
f_D&=&204(11)~{\rm MeV}~,\nnb\\
\hat m_c&=&1490(77)\lrar\overline{m}_c(\overline{m}_c)=1286(66)~{\rm MeV}~.
\label{eq:mc}
\eea 
which we consider as improvement of the result obtained from the same sum rule and at the same subtraction point  by \cite{SNFB2}\,\footnote{An extended discussion about the value of $f_D$ at  different subtraction points will be done in the next section.}:
\beq
f_D=205(20)~{\rm MeV}~,~~~~~\overline{m}_c(\overline{m}_c)= 1100 (40) ~{\rm MeV}~.
\eeq 
The smaller errors in the present analysis, come from more precise input parameters, more complete NP-corrections included into the OPE and more constrained range of $t_c$-values.The value obtained in Eq. (\ref{eq:mc}) also agrees within errors
with the accurate determination from charmonium systems quoted in Table~\ref{tab:param} though less accurate. The main sources of errors from the present determination can be found in Table \ref{tab:tau}. One can notice that the contributions of the $d=6$ condensates are negligible for $f_D$ (less than 0.3 MeV) and small for $m_c$ ($\la \bar dd\ra^2$ and $\la G^3\ra$ which contribute respectively to 17 and 6 MeV). 
{\scriptsize
\begin{table}[hbt]
\setlength{\tabcolsep}{0.12pc}
 \caption{
Central values and corresponding errors for $f_P$ and $\hat m_Q$ in units of MeV from the LSR at the subtraction point $\mu=\tau^{-1/2}$. We have used $\hat m_Q$ in Eq. (\ref{eq:heavymass}) for getting $f_P$. The +(resp. --) sign means that the values of $f_P,~\hat m_Q$  
increase (resp. decrease) when the input increases (resp. decreases). The relative change of sign from $c$ to $b$ in some errors is due to the effects of $\tau m_Q^2$ appearing the OPE. Notice that the error in $\la G^2\ra$ also affects the $\la G^3\ra$ contribution. The Total error comes from a quadratic sum.}
    {\small
\begin{tabular}{lccccccccccc}
&\\
\hline
&Value&$t_c$&$\alpha_s$&$\alpha_s^3$&$m_Q$&$\la\bar dd\ra$&$\la G^2\ra$&$M^2_0$& $\la\bar dd\ra^2$&$\la G^3\ra$& Total\\
\hline
$f_D$&204&+4&$-9$&+3&$-2$&+3.5&+0.5&$-0.5$&$-0.01$&+0.03&11\\
$f_B$&201&+7&$-10$&+1&$-2$&+1.9&+0.05&$-0.25$&$-0.00$&$+0.00$&13\\
\hline
$\hat m_c$&1457&$-44$&$-64$&$-24$&$0$&+22&+5&$-38$&+1.5&$-0.8$&93\\
$\hat m_b$&7272&$-150$&$-114$&$-14$&$0$&+20&+5&$-39$&-13&$-14$&195\\
\hline
\end{tabular}
}
\label{tab:tau}
\end{table}
}
\subsection*{\b Extension of the analysis to $f_B$ and $\hat m_b$}
\nin
\begin{figure}[hbt] 
\begin{center}
\centerline {\hspace*{-7.5cm} a) }\vspace{-0.6cm}
{\includegraphics[width=7cm]{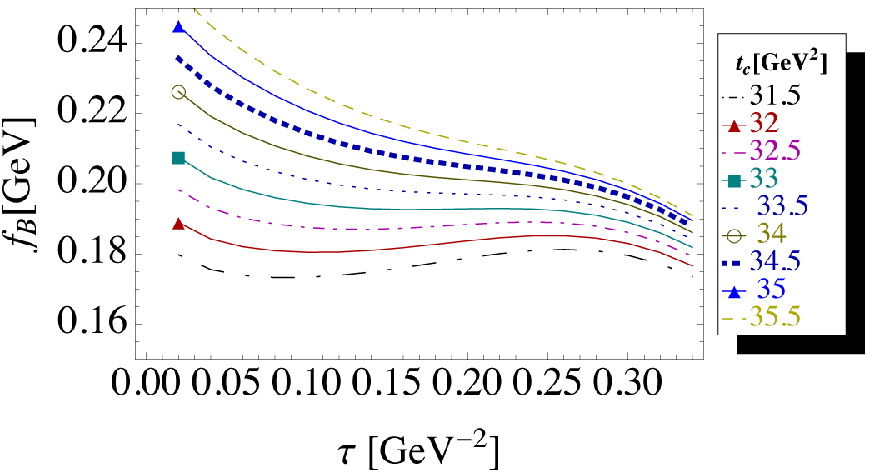}}
\centerline {\hspace*{-7.5cm} b) }\vspace{-0.3cm}
{\includegraphics[width=7cm]{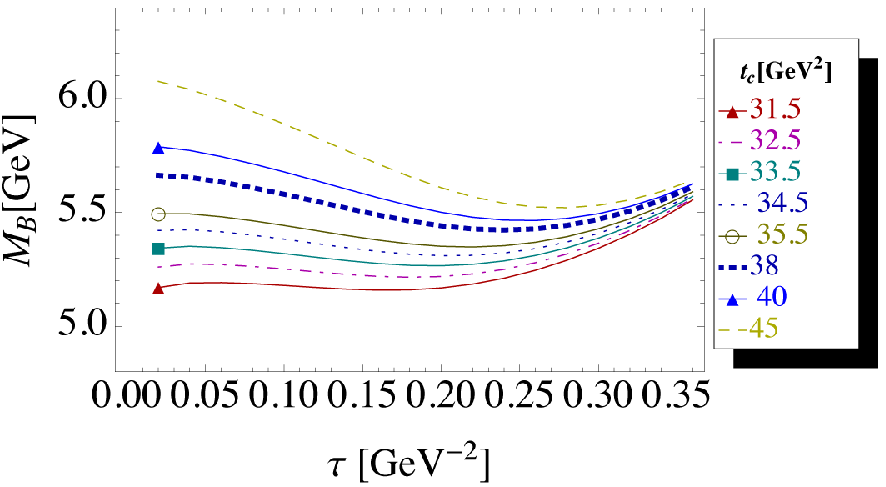}}
\caption{
\scriptsize 
{\bf a)} $\tau$-behaviour of $f_B$ from  ${\cal L}_{\bar db} $ for different values of $t_c$, for a given value of the subtraction point $\mu=\tau^{-1/2}$ and for $\hat m_b=7292$ MeV as given in Eq. (\ref{eq:heavymass}); {\bf b)} the same as a) but for $M_B$ from ${\cal R}_{\bar db}$. }
\label{fig:fbtau} 
\end{center}
\end{figure} 
\nin
We extend the previous analysis to the case of the $b$-quark. We show in Fig. (\ref{fig:fbtau}) the $\tau$-behaviours of $f_B$ and $M_B$ for different values of $t_c$.  
\begin{figure}[hbt] 
\begin{center}
\centerline {\hspace*{-7.5cm} a) }\vspace{-0.6cm}
{\includegraphics[width=7cm  ]{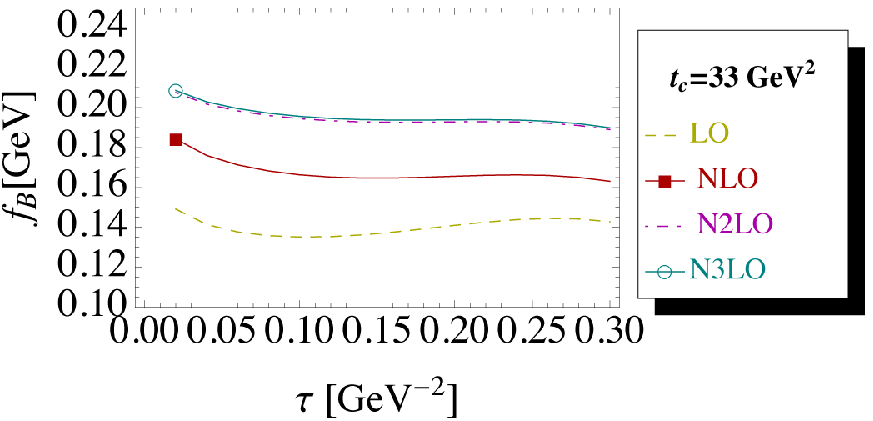}}
\centerline {\hspace*{-7.5cm} b) }\vspace{-0.3cm}
{\includegraphics[width=7cm  ]{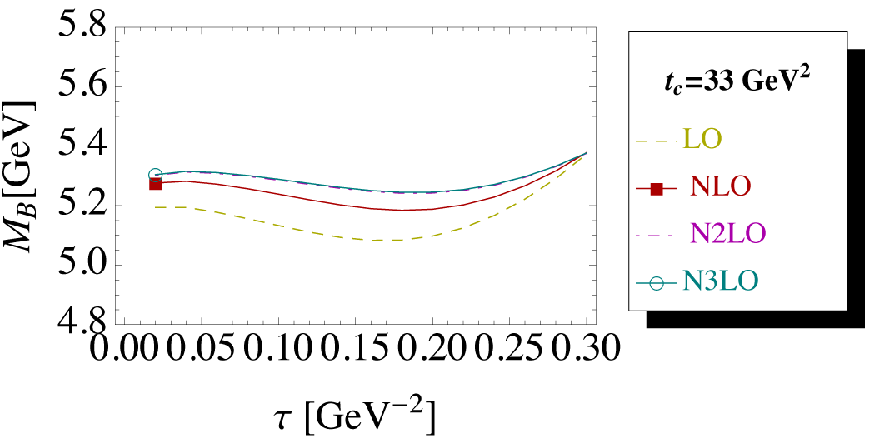}}
\caption{
\scriptsize 
$\tau$-behaviour of $f_B$ from LSR for $t^B_c$= 33 GeV$^2$, for $\hat m_b=7292$ MeV, for a given value of the subtraction point $\mu=\tau^{-1/2}$ GeV and for different truncations of the QCD PT series; {\bf b)} the same as a) but for $M_B$.
}
\label{fig:nlob} 
\end{center}
\end{figure} 
\nin
One can see, that in this channel, $\tau$-stability for $f_B$ is reached  at\,\footnote{ The apparent minima at $\tau\leq 0.1$ GeV$^2$ obtained for lower values of tc corresponds to the region where the continuum contribution to the spectral integral is dominant and should not be considered.}:
\beq
\tau^B_{0}\simeq (0.2\sim 0.26)~{\rm GeV}^{-2}~,~~~~~t^B_c\simeq (33\to 35) ~{\rm GeV}^2,
\label{eq:tcb}
\eeq
while, like in the case of $M_D$, $M_B$ stabilizes for a larger range of values\,\footnote{Like in the case of the charm quark, we shall not consider values of  $t_c\leq 32.5$ GeV$^2$ where the $\tau$-stabilty disappears, when one requires the sum rule to reproduce $M_B$.}:
\beq
t^B_c\simeq (33\to  45) ~{\rm GeV}^2~.
\label{eq:tcb2}
\eeq
We show in Fig. (\ref{fig:nlob}) the predicted values of $f_B$ and $M_B$ for a given value of $\hat m_b$ and for different truncations of the PT QCD series. 
Using the same procedure as in the charm quark case and considering the range of $t_c$ in Eq. (\ref{eq:tcb}), where the central values of $f_B$ and $\hat m_b$, in units of MeV in Table \ref{tab:tau} comes from an arithmetic average of different optimal values in the range of $t_c$ in Eq. (\ref{eq:tcb}), we deduce the estimate in units of MeV:
\bea
f_B&=&201(13)\nnb\\
\hat m_b&=&7272(195)~~\lrar~~\overline{m}_b(\overline{m}_b)=4164(112)~,
\eea
which we again consider as improvement of the result from \cite{SNFB2}:
\beq
f_B=203(23)~{\rm MeV},~~~~~~~~~
\overline{m}_b(\overline{m}_b)=4050(60)~{\rm MeV}~,
\eeq
obtained from the same sum rule. 
\begin{figure}[hbt] 
\begin{center}
\centerline {\hspace*{-7.5cm} a) }\vspace{-0.6cm}
{\includegraphics[width=7cm  ]{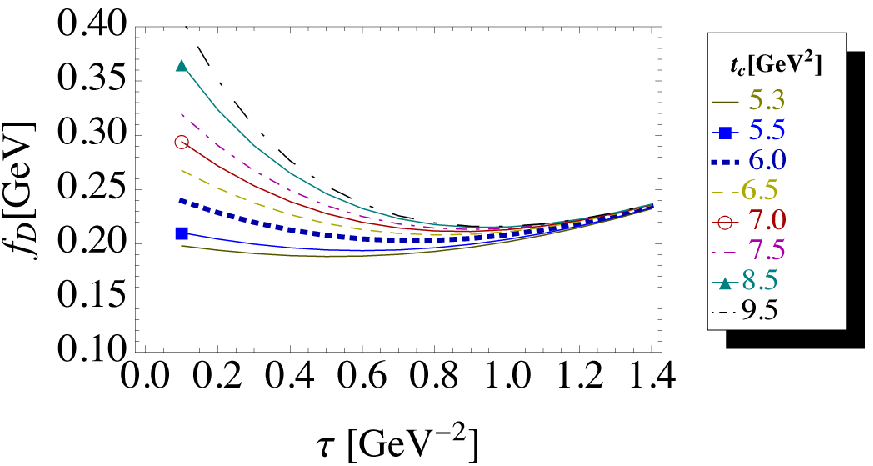}}
\centerline {\hspace*{-7.5cm} b) }\vspace{-0.3cm}
{\includegraphics[width=7cm  ]{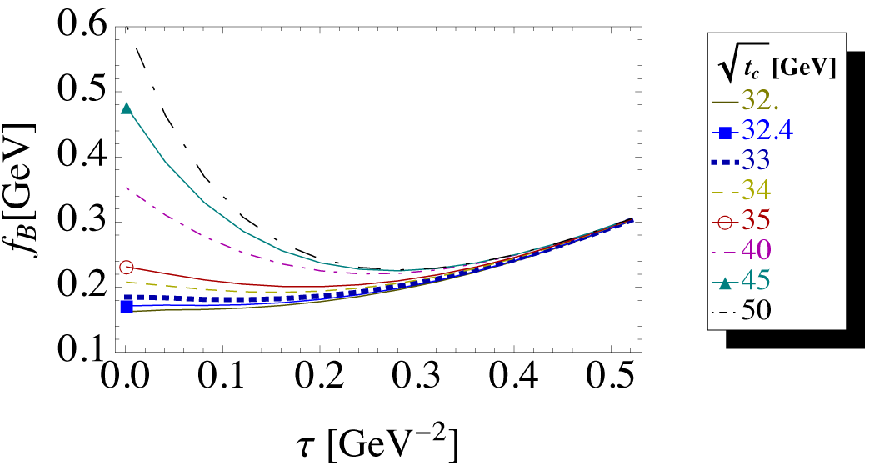}}
\caption{
\scriptsize 
{\bf a)} $\tau$-behaviour of $f_D$ from LSR for different values of $t_c$, for a given value of the subtraction point $\mu=1.4$ GeV and for $\hat m_c$=1467 MeV; {\bf b)} the same as a) but for $f_B$, using $\mu=3$ GeV and $\hat m_b$=7292 MeV}
\label{fig:fdbmu} 
\end{center}
\end{figure} 
\nin
\section{Effects of the subtraction point on  $f_{D,B}$ from LSR}
\begin{figure}[hbt] 
\begin{center}
{\includegraphics[width=7cm  ]{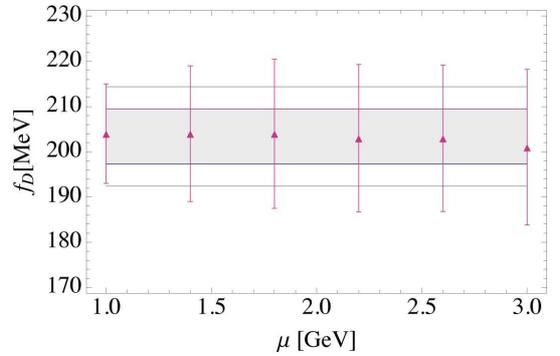}}
\caption{
\scriptsize 
Values of  $f_D$ from LSR at different values of the subtraction point $\mu$ and for $\hat m_c$=1467 MeV. The filled (grey) region is the average with the corresponding averaged errors. The
dashed horizontal lines are the values if one takes the errors from the best determination.}
\label{fig:fd_mean} 
\end{center}
\end{figure}

\nin
The choice of subtraction points is also one large source of errors and discrepancies
in the existing literature. In order to cure these weak points, we extract the values of $f_{D,B}$ and the corresponding errors  at a given value of the subtraction point $\mu$. We show in Fig. (\ref{fig:fdbmu}) the $\tau$-behaviour of $f_{D,B}$ for given values of $\mu$ and $\hat m_{c,b}$. We show in Tables \ref{tab:fd_mu} and \ref{tab:fb_mu}, the results of the analysis including the different sources of the errors,
 where the typical sizes normalized to the values of $f_{D,B}$ are:\\
-- $f_D$:
$(7\sim 8)$\% from $t_c$,  $(0.7\sim 2)$\% from the PT contributions, 0.5\% from $m_c$ and $(0.9\sim 1.6)$\% from the NP-contributions. \\
 --  $f_B$ are: $(2\sim 4)$\% from $t_c$, 4\% from the PT contributions, 4\% from $m_b$ and $(0.8\sim 1.5)$\% from the NP-contributions. \\
We show in Figs. (\ref{fig:fd_mean})  and  (\ref{fig:fb_mean}), the set of ``QSSR data points " obtained in this way for different values of $\mu$. 
\nin
{\scriptsize
\begin{table}[hbt]
\setlength{\tabcolsep}{0.18pc}
 \caption{
Central values and corresponding errors for $f_D$ in units of MeV from the LSR at different values of the subtraction point $\mu$ in units of MeV and for $\hat m_c$=1467 MeV. The +(resp. --) sign means that the values of $f_D$  
increase (resp. decrease) when the input increases (resp. decreases).  The Total error comes from a quadratic sum.}
    {\small
\begin{tabular}{cccccccccccc}
&\\
\hline
$\mu$&$f_D$&$t_c$&$\alpha_s$&$\alpha_s^3$&$m_c$&$\la\bar dd\ra$&$\la G^2\ra$&$M^2_0$& $\la\bar dd\ra^2$&$\la G^3\ra$& Total\\
\hline
1.4&204&+14&$-1.3$&+4&$-1$&+3&+1&$+0.6$&$+0.6$&+0.6&15.0\\
1.8&204&+16&$-1.2$&+2.7&$-0.9$&+2.3&+0.3&$+0.4$&0.0&0.0&16.5\\
2.2&203&+16&$-1.0$&+2.2&$-0.7$&+2.1&+0.3&$+0.3$&0.0&0.0&16.3\\
2.6&203&+16&$-1.1$&+1.5&$-1.1$&+1.6&+0.3&0.0&$-0.6$&$-0.5$&16.2\\
3.0&201&+17&$-0.8$&+1.2&$-0.8$&+1.6&+0.3&$+0.1$&$-0.5$&$-0.5$&17.2\\
\\
\hline
\end{tabular}
}
\label{tab:fd_mu}
\end{table}
}
{\scriptsize
\begin{table}[hbt]
\setlength{\tabcolsep}{0.16pc}
 \caption{
Central values and corresponding errors for $f_B$  in units of MeV from the LSR and MSR at different values of the subtraction point $\mu$ in units of GeV for $\hat m_b=7292$ MeV. The +(resp. --) sign means that the values of $f_B$  
increase (resp. decrease) when the input increases (resp. decreases).  The Total error comes from a quadratic sum.}
    {\small
\begin{tabular}{cccccccccccc}
&\\
\hline
$\mu$&$f_B$&$t_c$&$\alpha_s$&$\alpha_s^3$&$m_b$&$\la\bar dd\ra$&$\la G^2\ra$&$M^2_0$& $\la\bar dd\ra^2$&$\la G^3\ra$& \it Total\\
\hline
{\it LSR}\\
3&196&+22&$-8.0$&-1.6&$-1.1$&+1.9&+0.1&$+0.4$&$0.0$&0.0&23.6\\
4&210&+23&$-7.6$&-0.3&$-1.2$&+1.7&+0.1&$-0.2$&0.0&0.0&24.3\\
5&213&+24&$-7.6$&+0.4&$-1.2$&+1.5&+0.1&$0.0$&0.0&0.0&25.3\\
6&217&+24&$-7.3$&+0.1&$-1.2$&+1.6&+0.1&0.0&$0.0$&$0.0$&25.2\\
7&218&+21&$-7.1$&+0.5&$-1.0$&+1.5&+0.1&$0.0$&$0.0$&$0.0$&22.3\\
{\it MSR}\\
3&183&+7&$-16$&0&$-2.5$&+2&0&$-5$&$0$&0&18.4\\
4&199&+10&$-22$&0&$-3$&+3&0&$-9$&0&0&26.1\\
5&216&+11&$-19$&+1&$-3$&+4&0&$-13$&0&0&26.0\\
6&227&+17&$-21$&0&$-4$&+3&0&$-17$&$0$&$0$&32.3\\
7&235&+20&$-21$&+0.5&$-3$&+4&0&$-20$&$0$&$0$&35.6\\

\hline
\end{tabular}
}
\label{tab:fb_mu}
\end{table}
}
\subsection*{\b Final results for $f_{D}$ and $f_{B}$ from LSR}
\nin
Using the fact that the ``physical observable" is independent of $\mu$,
 we average (fit horizontally) the different data points of $f_D$  from LSR in Tables \ref{tab:tau} and \ref{tab:fd_mu} and Fig.  (\ref{fig:fd_mean}) (red triangle). The average is represented by the horizontal band in Fig. (\ref{fig:fd_mean}).
The narrower (grey) domain corresponds to the resulting averaged error, while the larger one corresponds to the case where the error from the most precise determination has been taken. A similar analysis for $f_B$ from LSR has been done using the data in Tables \ref{tab:tau} and \ref{tab:fb_mu} and Fig.  (\ref{fig:fb_mean}).
We deduce from this analysis, the final  results:
\bea
f_D&=&204(6)~{\rm MeV} \equiv  1.56(5) f_\pi~\nnb\\
f_B\vert_{LSR} &=& 207(8)~{\rm MeV} \equiv  1.59(6) f_\pi~,
\label{eq:fd_final}
\eea
where the quoted errors are the averaged errors. The previous errors are multiplied by about 1.8 for $f_D$ and 1.65 for $f_B$ if one keeps the errors from the most precise determinations. 
\subsection*{\b Final value of  $\hat m_b$ from LSR}
\nin
\begin{figure}[hbt] 
\begin{center}
{\includegraphics[width=7cm  ]{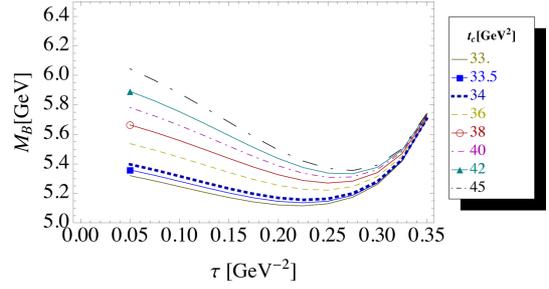}}
\caption{
\scriptsize 
 $\tau$ behaviour  of  $M_B$ from  LSR  for different values of $t_c$,   for $\hat m_b$=7292 MeV and at the subtraction point $\mu=M_b$. }
\label{fig:mb_M_lsr} 
\end{center}
\end{figure} 
\nin
In addition to the sum rule ${\cal R}_{\bar db}$ subtracted at $\mu=\tau^{-1/2}$, we also notice that the sum rule ${\cal R}_{\bar db}$ subtracted at $\mu=M_b$, where the $\log{(\mu/M_b)}$-term disappears in the QCD expression,  presents $\tau$-stability [see Fig. (\ref{fig:mb_M_lsr})] and can then provide another estimate of $\hat m_b$. The result is given in Table \ref{tab:mb_mu}. Taking the average of this result with the previous one in Table \ref{tab:tau}, we deduce in units of MeV:
\beq
\hat m_b\vert_{LSR}=7326(178)~~\lrar~~\overline{m}_b(\overline{m}_b)\vert_{LSR}=4195(102)~,
\eeq
{\scriptsize
\begin{table}[hbt]
\setlength{\tabcolsep}{0.25pc}
 \caption{
Central values and corresponding errors for $\hat m_b$ in units of MeV  from  LSR and MSR at different values of the subtraction point $\mu$ in units of GeV. The +(resp. --) sign means that the values of $\hat m_b$  
increase (resp. decrease) when the input increases (resp. decreases).  The Total error comes from a quadratic sum.}
    {\small
\begin{tabular}{ccccccccccc}
&\\
\hline
$\mu$&$\hat m_b$&$t_c$&$\alpha_s$&$\alpha_s^3$&$\la\bar dd\ra$&$\la G^2\ra$&$M^2_0$& $\la\bar dd\ra^2$&$\la G^3\ra$& \it Total\\
\hline
{\it LSR}\\
$M_b$&7586&-419&$-95$&-4&$+7$&+2&-26&$0$&$0$&431\\
{\it MSR}\\
3&7188&-295&$-110$&-6&$+6$&+1&-174&$+5$&$-0.5$&360\\
4&7360&-301&$-102$&-6&$+5$&+1&-178&$+4$&$-1$&365\\
5&7490&-306&$-99$&-4&$+8$&+1&-179&$+5$&$0$&368\\
6&7598&-310&$-99$&-4&$+9$&+1&-179&$+5$&$0$&372\\
7&7686&-312&$-97$&-4&$+9$&+1&-180&$+4$&$-1$&374\\
\hline
\end{tabular}
}
\label{tab:mb_mu}
\end{table}
}
\section{$Q^2=0$ moment  sum rules (MSR) for the $B$ meson}
\subsection*{\b Convergence of the PT series}
\begin{figure}[hbt] 
\begin{center}
\centerline {\hspace*{-7.5cm} a) }\vspace{-0.6cm}
{\includegraphics[width=7cm  ]{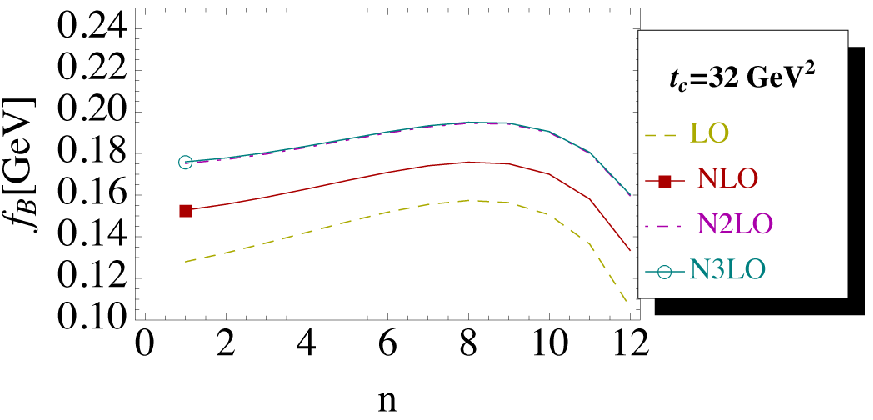}}
\centerline {\hspace*{-7.5cm} b) }\vspace{-0.3cm}
{\includegraphics[width=7cm  ]{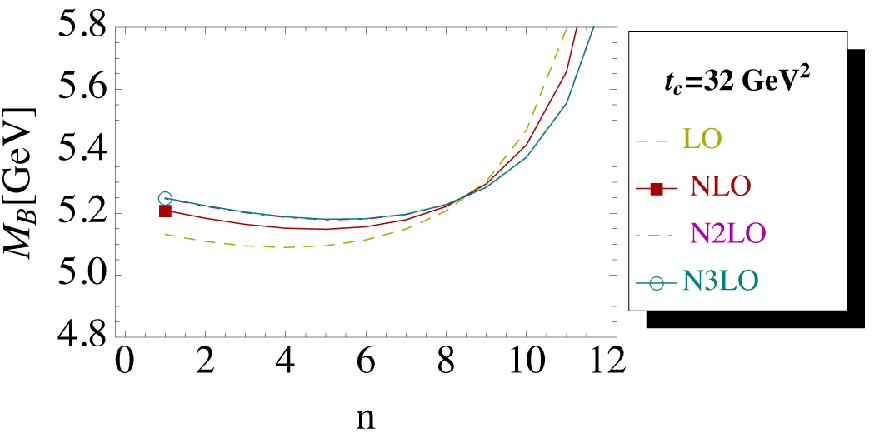}}
\caption{
\scriptsize 
{\bf a)} $n$-behaviour of $f_B$ from MSR for $t^B_c$= 32 GeV$^2$, for $\hat m_b=7292$ MeV, for a given value of the subtraction point $\mu$=4 GeV and for different truncations of the QCD PT series; {\bf b)} the same as a) but for $M_B$.
}
\label{fig:nlo_momb} 
\end{center}
\end{figure}

\nin
We show in Fig. (\ref{fig:nlo_momb}) the $n$-behaviours of $f_B$ and $M_B$ for different values of $t_c$, where one can realize a good convergence when the N3LO term is included.The convergence of the PT series is comparable with the one of LSR shown in Fig. (\ref{fig:nlob}).
\nin
\begin{figure}[hbt] 
\begin{center}
\centerline {\hspace*{-7.5cm} a) }\vspace{-0.6cm}
{\includegraphics[width=7cm  ]{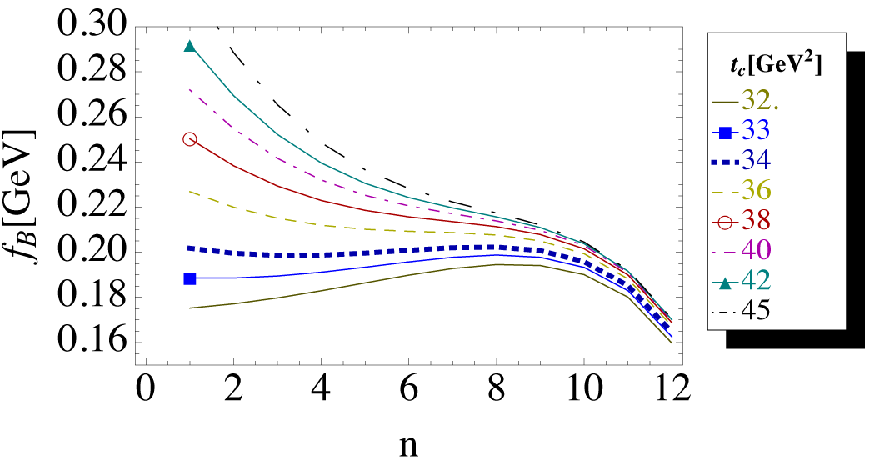}}
\centerline {\hspace*{-7.5cm} b) }\vspace{-0.3cm}
{\includegraphics[width=7cm  ]{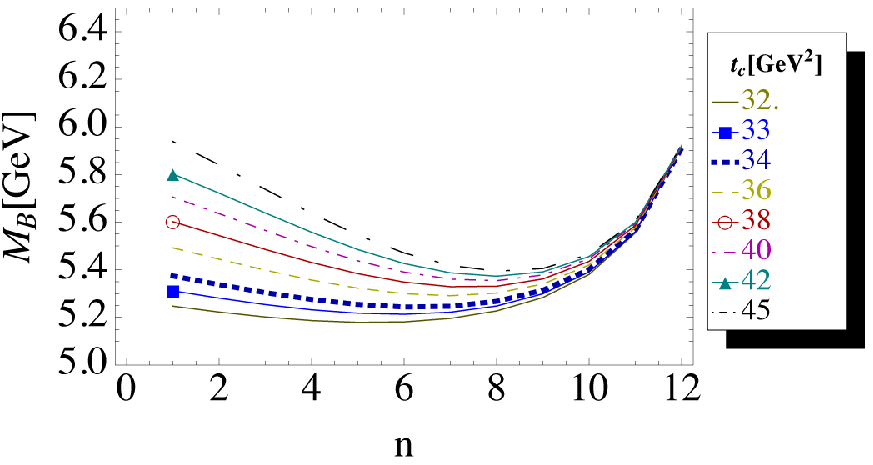}}
\caption{
\scriptsize 
{\bf a)} $n$ behaviour  of  $f_B$ from  MSR  for different values of $t_c$, for $\hat m_b$=7292 MeV and at the subtraction point $\mu=4$ GeV; {\bf b)} the same as a) but for $M_B$. }
\label{fig:fb_n} 
\end{center}
\end{figure} 
\nin
\subsection*{\b Optimal values of $f_B$ and $\hat m_b$ from MSR}
\begin{figure}[hbt] 
\begin{center}
{\includegraphics[width=7cm  ]{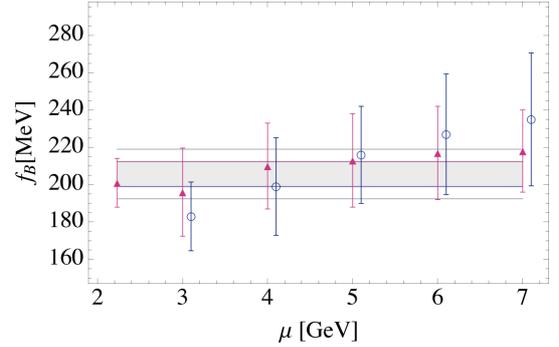}}
\caption{
\scriptsize 
{\bf } Values of  $f_B$ from LSR (red triangle) and from MSR (blue open circle) at different values of the subtraction point $\mu$ and for $\hat m_b$=7292 MeV. The filled (grey) region is the average with the corresponding averaged errors. The
dashed horizontal lines are the values if one takes the errors from the best determination.}
\label{fig:fb_mean} 
\end{center}
\end{figure}

\nin
Using a similar procedure as for the LSR, we study, in the case of MSR, the $n$-and $t_c$-stabilities of 
$f_B$ and $\hat m_b$ for different values of the subtraction point $\mu$. The analysis is illustrated in Fig. (\ref{fig:fb_n}). The results are shown in Tables \ref{tab:fb_mu} and \ref{tab:mb_mu}. 
One can notice that the sum rule does not stabilize for $\mu < 2$ GeV, while for other values of $\mu$, the range of values $t_c$ at which the $n$-stability is reached depends on the value of the subtraction point $\mu$ and are inside the range 32--42 GeV$^2$. 
 We show the results in Table \ref{tab:fb_mu} an in Fig. (\ref{fig:fb_mean}) from which we deduce the result from the moments in units of MeV:
\bea
f_B\vert_{MSR}&=&203(11)~,\nnb\\
\hat m_b\vert_{MSR}&=&7460(164)~\lrar~\overline{m}_b(\overline{m}_b)\vert_{MSR}=4272(94)~,\nnb\\
\eea
\nin
\begin{figure}[hbt] 
\begin{center}
{\includegraphics[width=7cm  ]{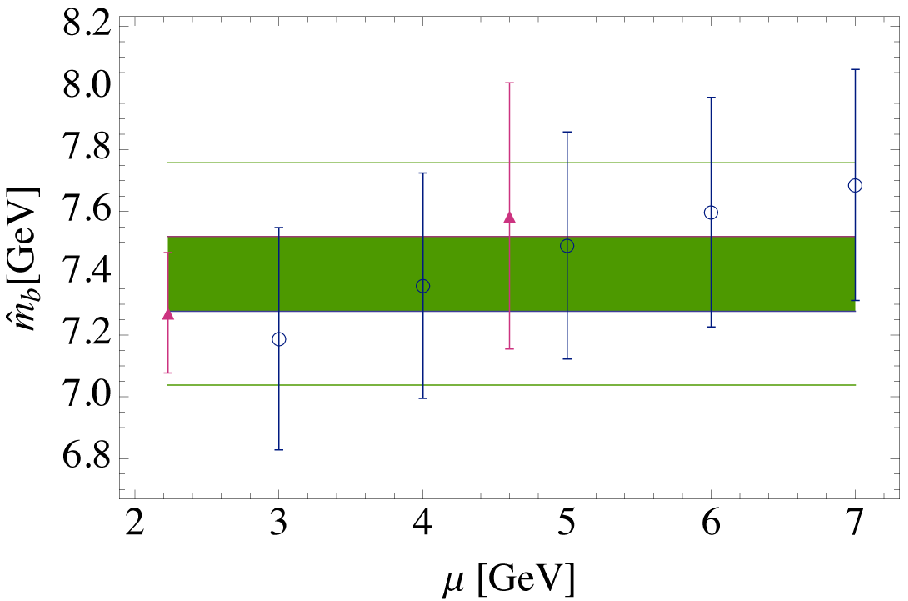}}
\caption{
\scriptsize 
{\bf } Values of  $\hat m_b$ from LSR (red triangle) and from MSR (blue open circle) at different values of the subtraction point $\mu$. Same caption as in Fig. \ref{fig:fb_mean}.}
\label{fig:mb_mean} 
\end{center}
\end{figure} 
\nin
\begin{figure}[hbt] 
\begin{center}
{\includegraphics[width=7cm  ]{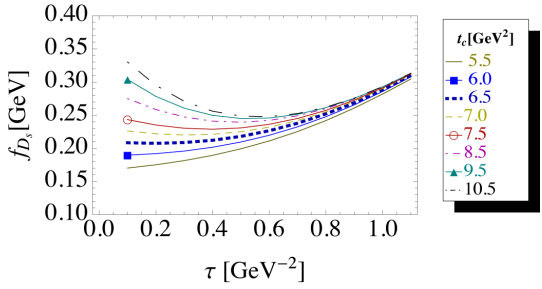}}
\caption{
\scriptsize 
$\tau$-behaviour of $f_{D_s}$ from  ${\cal L}_{\bar dc} $ for different values of $t_c$, for a given value of the subtraction point $\mu=1.4$ GeV and for $\hat m_c=1467$ MeV as given in Eq. (\ref{eq:heavymass}). }
\label{fig:fds_mu14} 
\end{center}
\end{figure} 
\nin
\begin{figure}[hbt] 
\begin{center}
{\includegraphics[width=7cm  ]{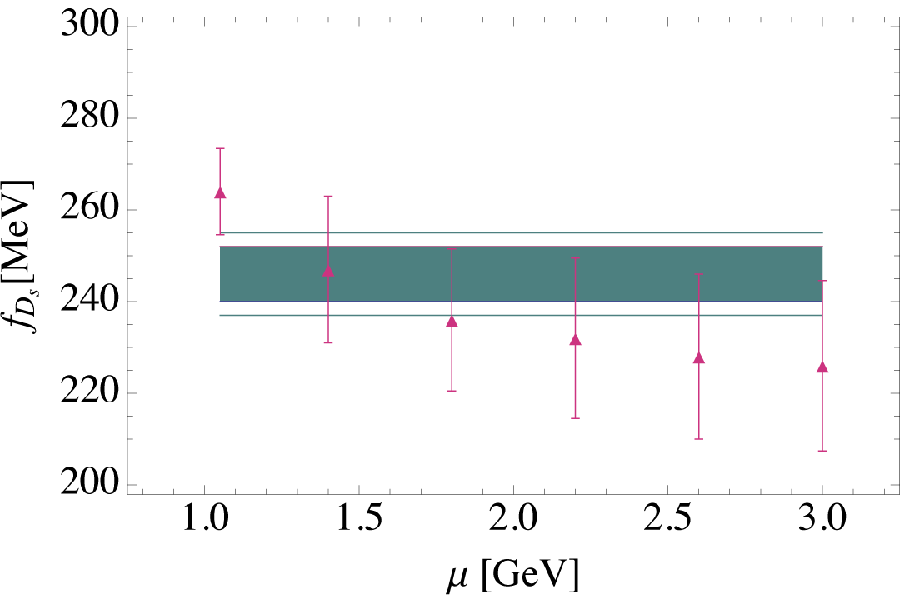}}
\caption{
\scriptsize 
{\bf}Values of  $f_{D_s}$ from LSR at different values of the subtraction point $\mu$ and for $\hat m_c$=1467 MeV. The filled (dark blue) region is the average with the corresponding averaged errors. The horizontal lines are the values if one takes the errors from the best determination.} 
\label{fig:fds_mean} 
\end{center}
\end{figure} 
\nin
\begin{figure}[hbt] 
\begin{center}
{\includegraphics[width=7cm  ]{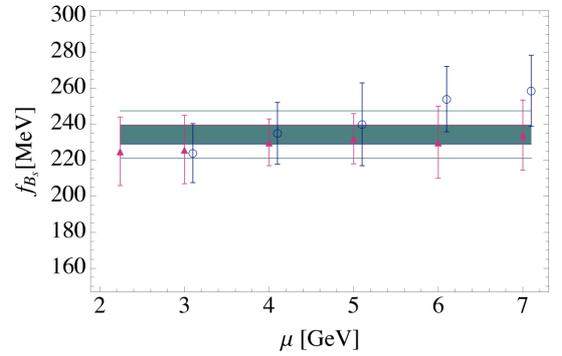}}
\caption{
\scriptsize 
{\bf } Values of  $f_{B_s}$ from LSR (red triangle) and from MSR (blue open circle) at different values of the subtraction point $\mu$ and for $\hat m_b$=7292 MeV. Same caption as in Fig. \ref{fig:fds_mean}.}
\label{fig:fbs_mean} 
\end{center}
\end{figure} 
\nin
{\scriptsize
\begin{table}[hbt]
\setlength{\tabcolsep}{0.15pc}
 \caption{
Central values and corresponding errors for $f_{D_s}$  from the LSR at different values of the subtraction point $\mu$ and for $\hat m_c$=1467 MeV. The +(resp. --) sign means that the values of $f_P,~\hat m_c$  
increase (resp. decrease) when the input increases (resp. decreases).  The Total error comes from a quadratic sum.}
    {\small
\begin{tabular}{cccccccccccc}
&\\
\hline
$\mu$&$f_{D_s}$&$t_c$&$\alpha_s$&$\alpha_s^3$&$m_c$&$\la\bar dd\ra$&$\la G^2\ra$&$M^2_0$&$m_s$& $\la\bar ss\ra$& Total\\
\hline
${\tau}^{-1/2}$&264&+8.2&$+2.8$&+3.2&$+0.2$&+0.8&+0.2&$+0.3$&$+1.2$&+0.5&9.4\\
1.4&247&+15&$+1.0$&+4.6&$+0.4$&+1.5&+0.7&$+0.9$&$+1.7$&+1.2&16\\
1.8&236&+15&$+1.1$&+3.3&$+1.1$&+1.0&+0.0&$+0.3$&+1.1&+0.5&15.5\\
2.2&232&+16.5&$+1.2$&+3.5&$+1.1$&+2.0&+1.4&$+1.5$&+2.2&+1.7&17.5\\
2.6&229&+17.6&$+0.3$&+1.8&$+0.1$&+1.2&+0.2&+0.5&$+1.1$&$+0.7$&18\\
3.0&226&+18.4&$+1.2$&+1.6&$+0.2$&+0.7&+0.1&$+0.2$&$+1.0$&$+0.4$&18.6\\
\\
\hline
\end{tabular}
}
\label{tab:fds_mu}
\end{table}
}
\begin{figure}[hbt] 
\begin{center}
\centerline {\hspace*{-7.5cm} a) }\vspace{-0.6cm}
{\includegraphics[width=7cm  ]{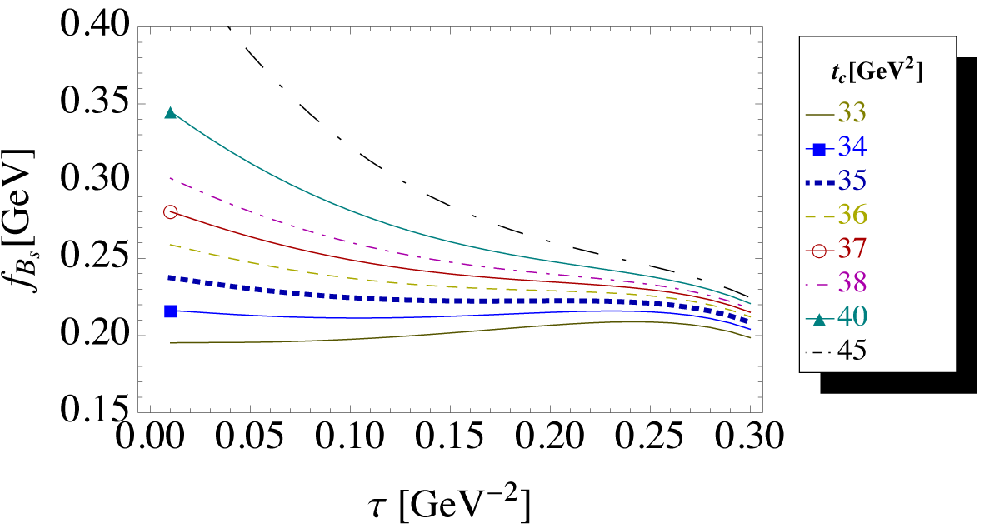}}
\centerline {\hspace*{-7.5cm} b) }\vspace{-0.3cm}
{\includegraphics[width=7cm  ]{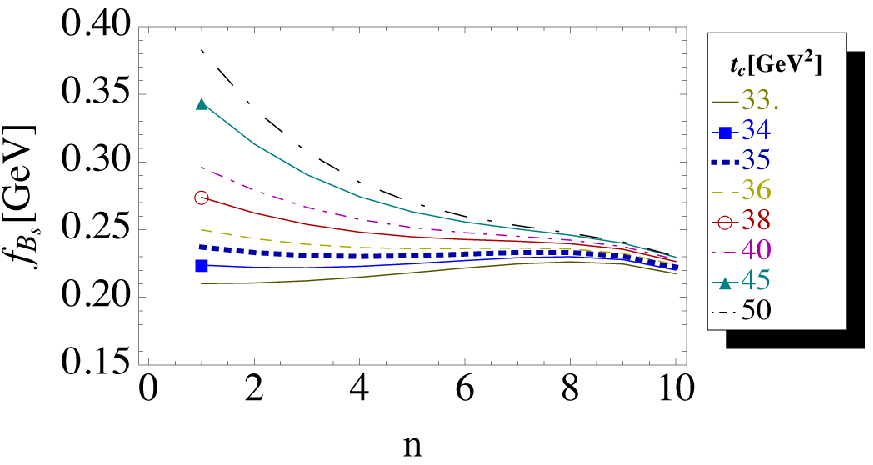}}
\caption{
\scriptsize 
{\bf a)} $\tau$ behaviour  of  $f_{B_s}$ from  LSR  for different values of $t_c$, for $\hat m_b$=7292 MeV and at the subtraction point $\mu=4$ GeV; {\bf b)} the same as a) but $n$ behaviour  of  $f_{B_s}$ from  MSR. }
\label{fig:fbs_mu4} 
\end{center}
\end{figure} 
\nin
{\scriptsize
\begin{table}[hbt]
\setlength{\tabcolsep}{0.14pc}
 \caption{
Central values and corresponding errors for $f_B$ in units of MeV from the LSR and MSR at different values of the subtraction point $\mu$ in units of MeV for $\hat m_b=7292$ MeV. The +(resp. --) sign means that the values of $f_B$  
increase (resp. decrease) when the input increases (resp. decreases).  The Total error comes from a quadratic sum.}
    {\small
\begin{tabular}{cccccccccccc}
&\\
\hline
$\mu$&$f_{B_s}$&$t_c$&$\alpha_s$&$\alpha_s^3$&$m_b$&$\la\bar dd\ra$&$\la G^2\ra$&$M^2_0$&$m_s$& $\la\bar ss\ra$& \it Total\\
\hline
{\it LSR}\\
$\tau^{-1/2}$&225&+18&$-1.9$&+3.5&$-1.3$&+1.4&+0.1&$-0.4$&$+0.4$&+0.7&19\\
3&226&+17&$-8.5$&+1.1&$-1.3$&+1.1&+0.0&$-0.1$&$+0.5$&+0.6&19\\
4&230&+10&$-8$&+0.8&$-1.3$&+1.1&+0.0&$0.0$&+0.5&+2.9&13\\
5&232&+11&$-8.3$&+0.7&$-1.9$&+1.2&$-0.3$&$-0.5$&+0.3&+2.5&14\\
6&230&+16&$-10.9$&+0.8&$-0.9$&+0.6&$-0.2$&-0.4&$+0.4$&$+2.5$&20\\
7&234&+16&$-10.6$&+0.7&$-1.2$&+1.2&+0.1&$-0.3$&$+0.4$&$+2.7$&20\\
{\it MSR}\\
3&224&+8.0&$-14$&+0.4&$-2.0$&+1.4&-0.2&$-1.1$&+1.0&+1.0&16\\
4&235&+13&$-10.6$&+1.0&$-2.1$&+1.3&$-0.1$&$-0.7$&+1.0&+1.2&17\\
5&240&+12.4&$-19.2$&+1.1&$-1.8$&+1.2&0.0&$-0.3$&+1.1&+0.7&23\\
6&254&+12&$-12.9$&+1.0&$-2.0$&+2.3&$-0.8$&$-2.4$&$+0.6$&+1.5&18\\
7&258.6&+14&$-13.2$&+1.0&$-2.4$&+2.4&$-0.5$&$-2.2$&$+0.7$&$+1.5$&20\\

\hline
\end{tabular}
}
\label{tab:fbs_mu}
\end{table}
}

\section{Final values of $f_{D},~f_{B}$ and $\hat m_{c,b}$}
As a final result of the present analysis, we take the average of the results from LSR for $f_D$ and $\hat m_c$. This result is given in Eq. (\ref{eq:mc}. The final results for $f_B$ and $\hat m_b$ come from the average of the ones from LSR and MSR shown in Figs. (\ref{fig:fb_mean}) and (\ref{fig:mb_mean}) , which are:
 \bea
f_B&=&206(7)~{\rm MeV}\equiv  1.58(5) f_\pi~,\nnb\\
\hat m_b&=&7398(121)~\lrar~\overline{m}_b(\overline{m}_b)=4236(69)~{\rm MeV}~,
\label{eq:fbmb_final}
\eea
where we have used the more precise value of $\overline{m}_b(\overline{m}_b)$ given in Table \ref{tab:param} for
getting $f_B$. 
One can notice that $f_D\simeq f_B$, confirming previous results quoted in Eq. (\ref{eq:fdb}). This (almost) equality instead of the $1/\sqrt{m_b}$ behaviour expected from HQET has been qualitatively interpreted in \cite{ZAL} using semi-local duality, while in \cite{SNFB4} large mass corrections to the HQET lowest order expression have been  found. These results are also confirmed by recent lattice calculations (see Table \ref{tab:result}). 
\begin{figure}[hbt] 
\begin{center}
\centerline {\hspace*{-7.5cm} a) }\vspace{-0.6cm}
{\includegraphics[width=7cm  ]{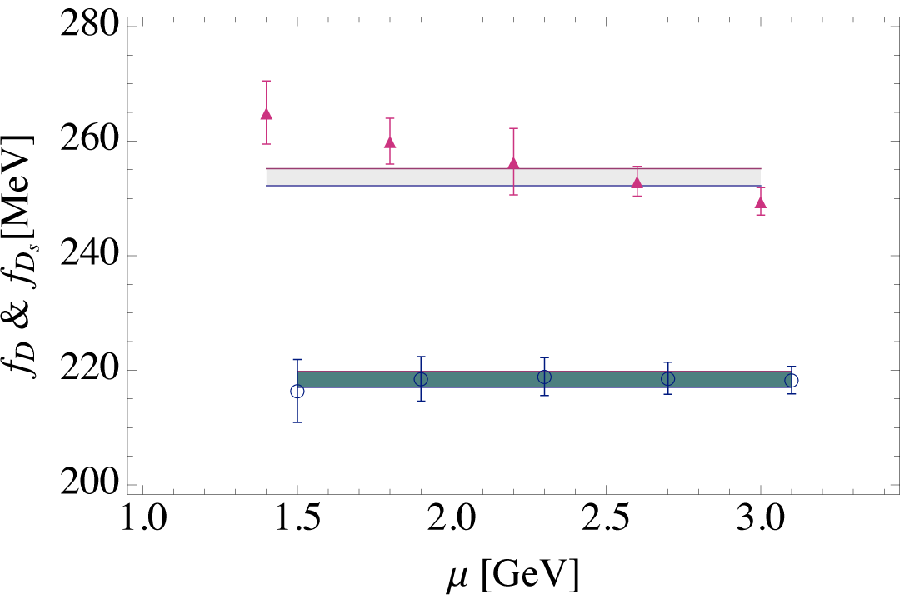}}
\centerline {\hspace*{-7.5cm} b) }\vspace{-0.3cm}
{\includegraphics[width=7cm  ]{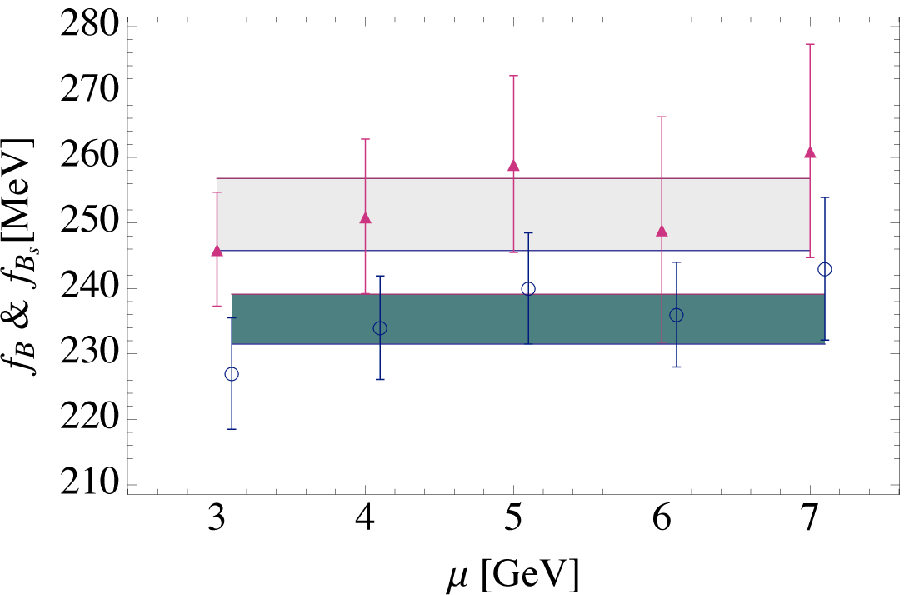}}
\caption{
\scriptsize 
{\bf a) } Upper bounds  of  $f_{D_s}$  (red triangle) and of   $f_{D}$ (blue open circle) at different values of the subtraction point $\mu$ and for $\hat m_c$=1467 MeV. The filled horizontal band is the average within the averaged error; {\bf b)} the same as a) but for $f_{B_s}$ and $f_{B}$ with $\hat m_b$=7292 MeV.}
\label{fig:fdfds_bound} 
\end{center}
\end{figure} 
\nin
\section{$SU(3)$ breaking and estimates of $f_{D_s}$ and $f_{B_s}$}
\nin
We extend the previous analysis for extracting $f_{D_s,B_s}$ by including the
$m_s$-corrections and by taking into account the $SU(3)$ breaking of the quark
condensate $\la\bar ss\ra/\la\bar dd\ra$ given in Table \ref{tab:param}. 
\subsection*{\b $f_{D_s}$ from LSR}
\nin
In so doing, we use the complete PT expression in $m_s$ of the QCD spectral function given
to order $\alpha_s$ by \cite{BROAD}. The massless expressions for N2LO and N3LO have been used. The non-perturbative contributions come from the expressions given by \cite{GENERALIS,JAMIN,JAMIN3} where we have taken into account corrections of ${\cal O}(m_s^2)$ for the $d=4$ condensates contributions while we have neglected the $m_s$ corrections for the $d=6$ condensates.  We show in Fig. (\ref{fig:fds_mu14}) the $\tau$-behaviour of $f_{D_s}$ for different values of $t_c$ at given $\mu=1.4$ GeV. The results for different values of $\mu$ are given in Table \ref{tab:fds_mu} and Fig (\ref{fig:fds_mean}). 
\subsection*{\b $f_{B_s}$ from LSR and MSR}
\nin
In this case, we only use the PT expression  to order $\alpha_s$ of the QCD spectral function  expanded up to order $m_s^2$ which is given by \cite{JAMIN3}.
The non-perturbative contributions are the same as in the case of $f_{D_s}$.  We show in Fig. (\ref{fig:fbs_mu4}) the $\tau$-behaviour and $n$-behaviour of $f_{B_s}$ from LSR and MSR for different values of $t_c$ at given $\mu=4$ GeV. The results for different values of $\mu$ are given in Table \ref{tab:fbs_mu} and Fig. (\ref{fig:fbs_mean}). 
\subsection*{\b Results for $f_{D_s}$ and $f_{B_s}$}
\nin
From the previous analysis , we deduce:
\bea
f_{D_s}&=&246(6)~{\rm MeV}\equiv 1.59(5)f_K\nnb\\
f_{B_s}&=&234(5)~{\rm MeV}\equiv 1.51(4)f_K
\label{eq:fdbs}
\eea
which, with the help of the results in Eqs. (\ref{eq:fd_final}) and (\ref{eq:fbmb_final}) lead to:
\beq
{f_{D_s}\over f_D}=1.21(4)~,~~~~~~~~{f_{B_s}\over f_B}=1.14(3)~.
\eeq
These results agree within the errors  with the ones obtained by using the semi-analytic 
expressions of the correlator to order $\alpha_s$~\cite{SNSU3}:
\beq
{f_{D_s}\over f_D}=1.15(4)~,~~~~~~~~{f_{B_s}\over f_B}=1.16(5)~,
\eeq
with data when available \cite{ROSNER,ASNER} and with recent lattice simulations (see Table \ref{tab:result}). 
\section{Rigourous model-independent upper bounds on $f_{D_{(s)},B_{(s)}}$}
Upper bounds on $f_D$ has been originally derived by NSV2Z \cite{NSV2Z} and improved four years later in \cite{SNZ,BROAD,GENERALIS} and more recently in \cite{KHOJ}. In this paper, we shall use LSR and the positivity of the continuum contributions to the spectral integral for obtaining the upper bounds on the decay constants. The procedure will be similar to the estimate done in previous sections where the optimal bound will be obtained at the minimum or inflexion point of the sum rules.  In the $D$ and $D_s$-meson channels, the LSR present a minimum which is well localized, while in the $B$ and $B_s$ channels, the LSR present instead an inflexion point which induces a new error for its localization, in addition to the errors induced by the QCD parameters which are the same as in the estimate of $f_{D,B}$ done in the previous sections. We show the results of the analysis for different values of the subtraction points in Fig. (\ref{fig:fdfds_bound})  from which we deduce the final results:
\bea
f_{D}&\leq &218.4(1.4)~{\rm MeV}\equiv 1.68(1)f_\pi\nnb\\
f_{B}&\leq&235.3(3.8)~{\rm MeV}\equiv 1.80(3)f_\pi
\label{eq:fdfb_bound}
\eea
and:
\bea
f_{D_s}&\leq &253.7(1.5)~{\rm MeV}\equiv 1.61(1)f_K\nnb\\
f_{B_s}&\leq&251.3(5.5)~{\rm MeV}\equiv 1.61(4)f_K
\label{eq:fdsfbs_bound}
\eea
These bounds are stronger  than earlier results in \cite{SNZ,BROAD,GENERALIS}, while the results for $f_{D,D_s}$ agree (within the large errors quoted there) with the ones in \cite{KHOJ}. These large errors come mainly from $m_c,\mu$ and $\bar \la dd\ra$. The previous bounds can be used for excluding some experimental data and some theoretical estimates. In deriving these bounds, we have only used the positivity of the spectral function and we have checked that the SVZ-expansion converges quite well both for the PT radiative and non-perturbative corrections such that the approximate series is expected to reproduce with a good precision the exact solution.
This fact can be (a posteriori) indicated by the remarkable agreement of our estimates with the lattice results. In this sense, we may state that the upper bound obtained previously is rigourous (at least within the SVZ framework). 
{\scriptsize
\begin{table}[hbt]
\setlength{\tabcolsep}{0.15pc}
 \caption{
Results from the open charm and beauty systems in units of MeV and comparison with experimental data and lattice simulations
using $n_f=2$ \cite{ETM,HEITGER} and $n_f=3$ \cite{HPQCD,MILC} dynamical quarks. $f_P$ are normalized as $f_\pi=130.4$ MeV. }
    {\small
\begin{tabular}{llll}
&\\
\hline
Charm&Bottom& Ref.    \\
\hline
$\overline{m}_c(\overline{m}_c)$&$\overline{m}_b(\overline{m}_b)$\\
1286(66) &4236(69)& This work \\
1280(40)&4290(140)&ETMC\cite{ETM}\\
$f_D$&$f_B$\\
$204(6)\equiv 1.56(5)f_\pi$&$206(7)\equiv 1.58(5)f_\pi$&This work\\
$\leq 218.4(1.4)\equiv 1.68(1)f_\pi$&$\leq 235.3(3.8)\equiv 1.80(3)f_\pi$& This work\\
207(9)&--& Data \cite{ROSNER,ASNER}\\
212(8)&195(12)& ETMC\cite{ETM}\\
--& 193(10)& ALPHA \cite{HEITGER} \\
207(4)&190(13)& HPQCD \cite{HPQCD}\\
219(11)&197(9)& FNAL \cite{MILC}\\
 $f_{D_s}$&$f_{B_s}$\\
$246(6)\equiv 1.59(5)f_K$&$234(5)\equiv 1.51(4)f_\pi$&This work\\
$\leq 253.7(1.5)\equiv 1.61(1)f_K$&$\leq 251.3(5.5)\equiv 1.61(4)f_K$& This work\\
260(5.4)&--& Data \cite{ROSNER,ASNER}\\
248(6)&232(12)& ETMC\cite{ETM}\\
--& 219(12)& ALPHA \cite{HEITGER}\\
248(2.5)&225(4)& HPQCD \cite{HPQCD}\\
260(11)&242(10)& FNAL \cite{MILC}\\

\hline
\end{tabular}
}
\label{tab:result}
\end{table}
}
\section{Summary and conclusions}
We have re-extracted the decay constants $f_{D,D_{s}}$ and $f_{B,B_{s}}$ and the running
quark masses $\overline{m}_{c,b}({m}_{c,b})$ using QCD
spectral sum rules (QSSR). We have used as inputs, the recent values of the QCD (non-)perturbative  parameters given in Table \ref{tab:param} and (for the first time) the renormalization group invariant quark and spontaneous masses in Eqs. (\ref{eq:heavymass}) and (\ref{eq:lightmass}). The results given in Eqs. (\ref{eq:fd_final}), (\ref{eq:fbmb_final}), (\ref{eq:fdfb_bound}) and (\ref{eq:fdsfbs_bound}) agree and improve existing QSSR results in the literature. Along the analysis, we have noticed that the values of the decay constants are very sensitive to the heavy quark mass and decrease when the heavy quark masses increase. Here we have used (for the first time) the scale independent Renormalization Group Invariant (RGI) heavy quark masses  in the analysis. We have translated the on-shell mass expressions of the PT spectral function known to N2LO into the $\overline{MS}$ one where (as has been already noticed in previous works \cite{SNFB2}) the PT series converge faster.  We have also remarked that $f_P$ and $m_Q$ are affected by the choice of the continuum threshold $t_c$ which gives the largest errors. Here, like in our previous works \cite{SNFB,SNFB4,SNFB2,SNFB3}, we have taken the conservative range of $t_c$-values where the $\tau$- or $n$-stability starts until the one where ones starts to have $t_c$-stability. We have also seen that the subtraction point $\mu$ affects the truncated results within the OPE which has been the sources of apparent discrepancies and large errors of the results in the literature. Here, we have considered carefully the results at each subtraction point and deduced, from these ``QSSR data", the final results  which should be independent on this arbitrary choice. In view of previous comments, we consider our results as improvements of the most recent ones to N2LO and using MDA in \cite{SNFB3,JAMIN3,KHOJ}.\\
The results on $f_D$ and $f_{D_s}$ agree within the errors with the data compiled in \cite{ROSNER,ASNER}, while the upper bound on $f_{D_s}$  can already exclude some existing data and theoretical estimates\,\footnote{Implications of the values of $f_{D,D_s}$ on the determinations of the CKM mixing angles have been discussed in details in \cite{SNFB8}.}. \\
As one can see in Table \ref{tab:result}, our results are comparable (in values and precisions) with recent lattice simulations including dynamical quarks \cite{ETM,HEITGER,HPQCD,MILC}\,\footnote{A summary  of the present results and comparisons with experiments and with the lattice ones will be presented elsewhere \cite{SNQCD12}.}. These agreements are not surprising as both methods start from the same observables (the two-point correlator though evaluated in two different space-times) and
use the 1st principles of QCD (here is the OPE in terms of the quarks and gluon condensates which semi-approximate the confinement regime). These agreements also confirm the accuracy of the MDA for describing the spectral function in the absence of a complete data, which has been tested earlier \cite{SNB1,SNB2} and in this paper from the charmonium and bottomium systems. MDA has been also successfully tested in the large $N_c$ limit of QCD in \cite{PERIS}. 


\begin{thebibliography}{99}
\bibitem{SNQCD12} S. Narison,  talk given at the 16th international QCD conference (QCD 12), 2-6th july 2012, Montpellier, arXiv:1209.2925 [hep-ph] (2012).

\bibitem{ROSNER}Review by J. Rosner and S. Stone in J. Beringer et al. (Particle Data Group), {\it Phys. Rev.} {\bf D86} (2012) 010001.


\bibitem{SVZ} M.A. Shifman, A.I. and Vainshtein and V.I. Zakharov,
{\it Nucl. Phys.} {\bf B147} (1979) 385; M.A. Shifman, A.I. and Vainshtein and V.I. Zakharov,
{\it Nucl. Phys.} {\bf B147} (1979) 448.

\bibitem{RRY} L.J. Reinders, H. Rubinstein and S. Yazaki, {\it Phys. Rept. }
{\bf 127}  (1985) 1. 

\bibitem{SNB1} S. 
Narison, {\it QCD as a theory of hadrons,
Cambridge Monogr. Part. Phys. Nucl. Phys. Cosmol.} {\bf 17} (2002) 1
[hep-h/0205006].

\bibitem{SNB2}S. Narison, {\it QCD
spectral sum rules ,  World Sci. Lect. Notes Phys.} {\bf 26}
(1989) 1.

\bibitem{SNB3}S. Narison, { Phys. Rept.} {\bf 84} (1982) 263; ibid, { Acta Phys. Pol.} {\bf B26} (1995) 687;  ibid, hep-ph/9510270 (1995).




\bibitem{NSV2Z}V.A. Novikov et al., 8th conf. physics and neutrino astrophysics (Neutrinos 78),
Purdue Univ. 28th April-2nd May 1978.

\bibitem{SNZ}S. Narison, {\it Z. Phys.} {\bf C14} (1982) 263. 


\bibitem{BROAD} D.J. Broadhurst and S.C. Generalis, Open Univ. report, OUT-4102-8/R (1982), unpublished.

\bibitem{GENERALIS}S.C. Generalis, Ph.D. thesis, Open Univ. report, OUT-4102-13 (1982), unpublished.

\bibitem{HQET} M.B. Voloshin and M.A. Shifman, {\it Sov.J. Nucl. Phys.} {\bf 45} (1987) 292; H.D. Politzer and M.B. Wise, {\it Phys. Lett.}  {\bf B206} (1988)  504,681. 

\bibitem{SNFB}S. Narison, {\it Phys. Lett.}  {\bf B198} (1987)  104; ibid, {\bf B285} (1992) 141.

\bibitem{SNFB4}S. Narison, {\it Phys. Lett.}  {\bf B279} (1992)  137;  
S. Narison, {\it Phys. Lett.}  {\bf B308} (1993)  365; S. Narison, {\it Z. Phys.} {\bf C55} (1992) 671. 

\bibitem{MARTI}G. Alexandrou et al., {\it Phys. Lett.}  {\bf B256} (1991)  60 and CERN-TH 6113 (1992);  C. Allton et al., {\it Nucl. Phys.} {\bf B349} (1991) 598; M. Lusignoli et al., Rome preprint 792 (1991); C. Bernard et al., Lattice workshop, Tallahassee (1990);

\bibitem{BROAD1}D.J. Broadhurst, {\it Phys. Lett.}  {\bf B101} (1981)  423 and private communication.

\bibitem{CHET}K.G. Chetyrkin and M. Steinhauser,  {\it Phys. Lett.}  {\bf B502} (2001)  104; hep-ph/0108017. 

\bibitem{JAMIN}M. Jamin and M. M\"unz, {\it Z. Phys.} {\bf C60} (1993) 569.

\bibitem{SNFB2} S. Narison,  {\it Phys. Lett.} {\bf B341} (1994) 73 ; S. Narison, {\it Nucl. Phys. Proc. Suppl.}  {\bf 74} (1999)  304.



\bibitem{TAR}R. Tarrach, {\it Nucl. Phys.} {\bf B183} (1981) 384.

\bibitem{COQUE} R. Coquereaux, {\it Annals of Physics} {\bf 125} (1980) 401; P. Binetruy and T. S\"ucker, {\it Nucl. Phys.} {\bf B178} (1981) 293; 

\bibitem{SNPOLE} S. Narison, {\it  Phys. Lett.} {\bf B197} (1987) 405; S. Narison, {\it  Phys. Lett.} {\bf B216} (1989) 191.

\bibitem{BROAD2} N. Gray, D.J. Broadhurst, W. Grafe, and K. Schilcher, {\it Z. Phys.} {\bf  C48} (1990) 673; 
J. Fleischer, F. Jegerlehner, O.V. Tarasov, and O.L. Veretin, {\it Nucl. Phys.} {\bf B539}
(1999) 671

\bibitem{CHET2}K.G. Chetyrkin and M. Steinhauser, {\it Nucl. Phys.} {\bf B573}
(2000) 617; K. Melnikov and T. van Ritbergen, hep-ph/9912391.
\bibitem{SNFB3}S. Narison, {\it Phys. Lett.}  {\bf B520} (2001)  115. 

\bibitem{JAMIN3}M. Jamin and B. O. Lange, {\it Phys. Rev.} {\bf D65} (2002) 056005.

\bibitem{KHOJ}A. Khodjamirian, {\it Phys. Rev.} {\bf D79} (2009) 031503.

\bibitem{PENIN} A. Penin and and M. Steinhauser, {\it Phys. Rev.} {\bf D65} (2002) 054006.

\bibitem{NEUBERT}D.J. Broadhurst and M. Grozin,  {\it Phys. Lett.}  {\bf B274} (1992)  421; E. Bagan,
P. Ball, V. Braun and H.G. Dosch,  {\it Phys. Lett.}  {\bf B278} (1992)  457; M. Neubert, {\it Phys. Rev.} {\bf D45} (1992) 2451;  V. Eletsky and A.V. Shuryak,  {\it Phys. Lett.}  {\bf B276} (1993)  365.  
\bibitem{NZ}S. Narison and V.I. Zakharov, {\it Phys. Lett.} {\bf B679} (2009) 355.

\bibitem{CNZ} K.G. Chetyrkin, S. Narison and V.I. Zakharov, {\it Nucl. Phys.} 
{\bf B550} (1999)  353;
S. Narison and V.I. Zakharov, {\it  Phys. Lett.} {\bf B522} (2001) 266.

\bibitem{ZAK} For reviews, see e.g.: V.I. Zakharov, {\it Nucl. Phys. Proc. Suppl.} 
{\bf 164} (2007) 240; S. Narison,  {\it Nucl. Phys. Proc. Suppl.} {\bf 164} 
 (2007) 225.

 \bibitem{FNR}E.G. Floratos, S. Narison and E. de Rafael, {\it Nucl. Phys.} 
{\bf B155} (1979) 155.

\bibitem{SNRAF}S. Narison and E. de Rafael, {\it Phys. Lett.} {\bf B103} (1981)57.

\bibitem{SNH10}S. Narison,  {\it Phys. Lett.} {\bf B693} (2010)  559; Erratum ibid 705 (2011) 544;
ibid, {\it Phys. Lett.} {\bf B706} (2011)  412; ibid, {\it Phys. Lett.} {\bf B707} (2012)  259. 

\bibitem{PERIS}E. de Rafael, {\it Nucl. Phys. Proc. Suppl.} {\bf 96} (2001) 316; S. Peris, B. Phily and E. de Rafael, {\it Phys. Rev. Lett.} {\bf 86} (2001) 14.

\bibitem{BELL}J.S. Bell and R.A. Bertlmann, {\it Nucl. Phys.} {\bf B227} (1983) 435;
R.A. Bertlmann, {\it Acta Phys. Austriaca} {\bf 53} (1981) 305;  R.A. Bertlmann 
and H. Neufeld, {\it Z. Phys.} {\bf C27} (1985)  437.

\bibitem{RUNDEC}K.G. Chetyrkin,  J.H. K\"uhn and M. Steinhauser, hep-ph/0004189 and references therein.

\bibitem{SNTAU}S. Narison, {\it Phys. Lett.} {\bf B673} (2009) 30.

\bibitem{BNP}E. Braaten, S. Narison and A. Pich, {\it Nucl. Phys.} {\bf B373}, 581 (1992); S. Narison and A. Pich, {\it Phys. Lett.} {\bf  B211} (1988) 183.

\bibitem{BETHKE} For a recent review, see e.g: S. Bethke, talk given at the 16th international QCD conference (QCD 12), 2-6th july 2012, Montpellier, arXiv:1210.0325 [hep-ex] (2012).

\bibitem{PDG} K. Nakamura et al. (PDG), {\it Journal Physics} {\bf G37}, 075021 (2010).

 \bibitem{SNmass} S. Narison, {\it  Phys.Rev.} {\bf D74} (2006) 034013.
 
 \bibitem{SNmass2}S. Narison,
arXiv:hep-ph/0202200; ibid, {\it Nucl.Phys.Proc.Suppl.} {\bf 86} (2000) 242; 
ibid, {\it Phys. Lett.} {\bf B216} (1989)  191; ibid, {\it Phys. Lett.} 
{\bf B358} (1995) 113; ibid, {\it Phys. Lett.} {\bf B466} (1999) 345; ibid, {\it Riv. Nuov. Cim.} {\bf 10N2} 
(1987) 1; S. Narison, H.G. 
Dosch, {\it Phys. Lett.} {\bf B417} (1998) 173; 
S. Narison, N. Paver, E. de Rafael and D. Treleani, {\it Nucl. Phys.} {\bf B212} 
 (1983) 365; S. Narison, E. de Rafael, {\it Phys. Lett.} {\bf B103} (1981) 57; C. 
Becchi, S. Narison, E. de Rafael, F.J. Yndurain, {\it Z. Phys.} {\bf C8} (1981)  335.

\bibitem{SNHmass}S. Narison, {\it Phys. Lett.} {\bf B197}(1987) 405 ; 
ibid, {\it Phys. Lett.} {\bf B341} (1994) 73 ; ibid, {\it Phys. Lett.} {\bf B520}  (2001) 115.

\bibitem{IOFFE} B.L. Ioffe and K.N. Zyablyuk, {\it Eur. Phys. J.} {\bf  C27}
(2003)  229 ; B.L. Ioffe, {\it Prog. Part. Nucl. Phys.} {\bf 56} (2006) 232.

\bibitem{Hbaryon}R.A. Albuquerque, S. Narison and M. Nielsen, {\it Phys. Lett.} {\bf B684} (2010)236.

\bibitem{JAMI2}Y. Chung et al.{\it Z. Phys.} {\bf C25} (1984)  151;  H.G. Dosch, 
Non-Perturbative Methods (Montpellier 1985);  
H.G. Dosch, M. Jamin and S. Narison, {\it Phys. Lett.} {\bf B220} (1989)  251.

\bibitem{HEID}B.L. Ioffe, {\it Nucl. Phys.} {\bf B188} (1981)  317; B.L. Ioffe, {\bf
B191} (1981) 591; A.A.Ovchinnikov and A.A.Pivovarov,
{\it Yad.\ Fiz.}  {\bf 48} (1988) 1135.
\bibitem{SNhl}S. Narison, Phys. Lett. {\bf B605} (2005) 319.

\bibitem{LNT}G. Launer, S. Narison and R. Tarrach, {\it  Z. Phys.} {\bf C26}
(1984) 433.
\bibitem{SNI}S. Narison, {\it Phys. Lett.} {\bf B300} (1993) 293; ibid {\bf B361} (1995) 121.
\bibitem{fesr} R.A. Bertlmann, G. Launer and E. de Rafael, 
{ Nucl. Phys.} {\bf B250} (1985) 61; R.A. Bertlmann et al., 
{ Z.\ Phys.}  {\bf C39} (1988) 231.
\bibitem{YNDU}F.J. Yndurain, hep-ph/9903457.
\bibitem{SNHeavy}S. Narison, {\it Phys. Lett.} {\bf B387} (1996) 162.

\bibitem{SNG} S. Narison, {\it Phys. Lett.} {\bf B361} (1995) 121;
S. Narison,  {\it Phys. Lett.} {\bf B624} (2005) 223.
\bibitem{NAVARRA} For reviews, see e.g.: F. S. Navarra, M. Nielsen, S. H. Lee, 
{\it Phys. Rep. } {\bf 497} (2010) 41; 
S. L. Zhu, {\it Int.  J.  Mod.  Phys.} {\bf  E 17}  (2008) 283.

\bibitem{STEELE} D. Harnett, R.T. Kleiv, T.G. Steele, H.-y. Jin, arXiv:1206.6776 [hep-ph];
ibid, arXiv:1208.3273 [hep-ph];
R. Berg, D. Harnett, R.T. Kleiv, T.G. Steele, {\it Phys.Rev.} {\bf D86} (2012) 034002; ibid, arXiv:1209.4102 [hep-ph].

\bibitem{NIELSEN}R.M. Albuquerque,  F. Fanomezana, S. Narison and A. Rabemananjara, {\it Phys. Lett.} {\bf B715} (2012) 129; ibid,  arXiv:1210.2990 [hep-ph]; 
R.D. Matheus, S. Narison, M. Nielsen, J.M. Richard, {\it Phys. Rev.} {\bf D75} (2007) 014005.

\bibitem{ZAL}S. Narison and K. Zalewski, {\it Phys. Lett.} {\bf B320} (1994) 369 .

\bibitem{SNSU3}S. Narison, {\it Phys. Lett.} {\bf B322} (1994) 247.

\bibitem{ASNER} D. Asner et al., Flavor averaging group, arXiv:1010.1589 [hep-ex], http://
www.slac.stanford.edu/xorg/hfag/charm/.

\bibitem{ETM} ETM collaboration: P. Dimopoulos et al., JHEP 1201 (2012) 046); N. Carrasco and A. Shindler, Lattice 2012 and  private communication from G. Rossi.

\bibitem{HEITGER} ALPHA collaboration: J. Heitger, talk given at the 16th international QCD conference (QCD 12), 2-6th july 2012, Montpellier.


\bibitem{HPQCD} HPQCD collaboration: C.T.H. Davies et al., arXiv:1008.4018 [hep-lat]. 

\bibitem{MILC} Fermilab and MILC collaboration: A. Bazavov et al., arXiv:1112.305 [hep-lat] (2011).

\bibitem{SNFB8}S. Narison, {\it Phys. Lett.} {\bf B668} (2008) 308.


\end{thebibliography}
\end{document}